\documentclass[12pt,preprint]{aastex}
\usepackage{multirow}
\usepackage{natbib}

\newcommand\arcpt{${{\lower3pt\hbox{$^{\prime\prime}$}}\atop{\raise4pt\hbox{.}}}$}
\newcommand\msun{$M_\odot$}

\slugcomment{submitted to the {\it Astronomical Journal}  2011 June 15}
\slugcomment{accepted 2011 July 11}

\shorttitle{AP~Col: The Closest PMS Star}
\shortauthors{Riedel et al.}

\begin{document}

\title{The Solar Neighborhood. XXVI. AP~Col: The Closest (8.4 pc)
Pre-Main-Sequence Star}

\author{Adric~R.~Riedel\altaffilmark{1}}

\affil {Department of Physics and Astronomy, Georgia State University,
Atlanta, GA 30302-4106} 

\email{riedel@chara.gsu.edu}

\author {Simon~J.~Murphy}

\affil {Research School of Astronomy and Astrophysics, The Australian
National University, Cotter Road, Weston Creek, ACT 2611, Australia}

\email{murphysj@mso.anu.edu.au}

\author {Todd~J.~Henry\altaffilmark{1}}

\affil {Department of Physics and Astronomy, Georgia State University,
Atlanta, GA 30302-4106}

\email {thenry@chara.gsu.edu}

\author{Carl~Melis}

\affil {Center for Astrophysics and Space Sciences, University of
California, San Diego, CA 92093-0424, USA.}

\email{cmelis@ucsd.edu}

\author{Wei-Chun~Jao\altaffilmark{1}}

\affil {Department of Physics and Astronomy, Georgia State University,
Atlanta, GA 30302-4106}

\email{jao@chara.gsu.edu}

\author{John~P.~Subasavage\altaffilmark{1}}

\affil {Cerro Tololo Inter-American Observatory, La Serena, Chile}

\email{jsubasavage@ctio.noao.edu}





\altaffiltext{1}{Visiting Astronomer, Cerro Tololo Inter-american
  Observatory.  CTIO is operated by AURA, Inc.\ under contract to the
  National Science Foundation.}

\begin{abstract}

We present the results of a multi-technique investigation of the
M4.5Ve flare star AP~Col, which we discover to be the nearest
pre-main-sequence star.  These include astrometric data from the CTIO
0.9m, from which we derive a proper motion of 342.0$\pm$0.5 mas
yr$^{-1}$, a trigonometric parallax of 119.21$\pm$0.98 mas
(8.39$\pm$0.07 pc), and photometry and photometric variability at
optical wavelengths.  We also provide spectroscopic data, including
radial velocity (22.4$\pm$0.3 km s$^{-1}$), lithium Equivalent Width
(EW) (0.28$\pm$0.02\AA), H$\alpha$ EW ($-$6.0 to $-$35\AA), {\it
vsini} (11 $\pm$ 1 km s$^{-1}$), and gravity indicators from the
Siding Spring 2.3-m WiFeS, Lick 3-m Hamilton echelle, and Keck-I HIRES
echelle spectrographs.  The combined observations demonstrate that
AP~Col is the closer of only two known systems within 10 pc of the Sun
younger than 100 Myr.  Given its space motion and apparent age of
12-50 Myr, AP~Col is likely a member of the recently proposed $\sim$40
Myr old Argus/IC~2391 association.

\end{abstract}

\section{Introduction}
\label{sec:intro}

For decades, young stars ($\lesssim$600 Myr) were thought only to
reside in giant star-forming regions like the Orion Nebula, the
Taurus-Auriga dark cloud, the Scorpius-Centaurus star forming region,
or other distant large molecular clouds.  Indeed, the only relatively
young stars near the Sun seemed to be in the Pleiades
(\citealt{Trumpler1921}; age $\sim$125 Myr, \citealt{Stauffer1998}), the
Sirius-Ursa Major moving group (\citealt{Eggen1958}; age $\sim$300
Myr, \citealt{Famey2008}) and the Hyades (\citealt{Eggen1958}; age
$\sim$600 Myr, \citealt{Famey2008}).  In the 1980s, this picture
changed as the stars TW Hydra and $\beta$~Pictoris were identified as
isolated T Tauri stars.  Eventually, both TW Hydra
\citep{de-la-Reza1989} and $\beta$~Pictoris
\citep{Barrado-y-Navascues1999} were found to be members of sparse
kinematic associations of pre-main-sequence stars under 100 Myr old,
and a number of other nearby associations
\citep{Zuckerman2004,Torres2008} were identified.

Currently, there are several reported associations less than 100 Myr
old with members closer than 100 pc, comprising associations such as 
$\sim8$ Myr old TW Hydra (now with over 30 members), $\sim12$ Myr
$\beta$~Pictoris, $\sim70$ Myr AB~Doradus, and $\sim40$ Myr
Argus/IC~2391.  These clusters, while sparse and not gravitationally
bound, are nearby easy-to-examine windows into the history and
processes of star formation, far younger and far more accessible than
classical open clusters like the Pleiades
\citep[133~pc,][]{Soderblom2005}.

The X-ray active M dwarf \object{AP Col} (=LP~949-015, LTT~2449,
SIPS~J0604-3433, 2MASS~J06045215-3433360) was identified as a UV-Ceti
type flare star as early as 1995 (\citealt{Ball1995}), whereupon it
was given its variable star designation.  It was studied by
\citet{Scholz2005} as one of three active M dwarfs detected within a
predicted distance of 8 pc, and again by \citet{Riaz2006}, where the
star's potential youth and proximity was again noted.  It was also
targeted for Lucky Imaging by \citet{Bergfors2010}, and for Speckle
Imaging with the USNO Specklecam on the CTIO 4-m by Mason \& Hartkopf
(2011, private communication).  The Research Consortium on Nearby Stars
(RECONS\footnote{www.recons.org}) searches for nearby stars in the
southern sky \citep{Henry2006} also identified this star as
potentially nearby, and it was put on the target list of the Cerro
Tololo Inter-american Parallax Investigation (CTIOPI) in 2004.  As
described in this work, the star was more recently investigated at the
Lick, Keck, and Siding Spring Observatories.

In conjunction with the new data presented here, we will first show
that the observed characteristics of AP~Col are signs of a youthful
age of less than 100 Myr and not interactions with a close companion.
We will then argue that its age and kinematics match those of the
Argus association defined by \citet{Torres2008}.

\section{Observations}

\subsection{Astrometry and Photometry}

All new astrometric and photometric observations were carried out at
the CTIO 0.9-m telescope, initially (1999-2003) under the aegis of the
National Optical Astronomy Observatory (NOAO) Surveys Program, and
later (2003-present) via the Small and Moderate Aperture Research
Telescope System (SMARTS) Consortium, as part of the long running
CTIOPI program \citep{Jao2005}.

The 158 $V_J$ filter observations of AP~Col used in our astrometric
and relative photometry sequences were obtained on 27 nights between
September 2004 and March 2011, utilizing the central 1024x1024 pixels
of the 0.9-m telescope's 2048x2046 Tek CCD with a 0.401\arcsec~pixel
scale and CTIO's $V_J, R_{KC}$, and $I_{KC}$\footnote{The central
wavelengths for $V_J, R_{KC}$, and $I_{KC}$ are 5475, 6425, and 8075
\AA, respectively.  The subscript ``J'' indicates Johnson, ``KC''
indicates Kron-Cousins (usually known as `Cousins'), and are
hereafter omitted.}  filters.  Additional details of the observing
protocols for the astrometry and photometry programs can be found in
\citet{Jao2005} and \citet{Winters2011}, respectively.

The same setup was used to obtain four nights of $VRI$ photometry,
interleaved with standard star observations from \citet{Graham1982},
\citet{Landolt1992}, and \citet{Landolt2007} at various airmasses.
The photometric observations were reduced via a custom IRAF pipeline
and transformed onto the Johnson-Kron-Cousins system, as described in
\citet{Henry2006}.

\subsection{Spectroscopy}

\subsubsection{CTIO 1.5-m RCspec}

To measure a spectral type, AP~Col was observed on the CTIO 1.5-m on
UT 14 March 2004 using the 32/I grating setup (6000-9600\AA, R=1500).
The resulting spectrum was reduced using standard IRAF procedures and
then classified as M4.5Ve using the ALLSTAR code \citep{Henry2002},
which compares it to spectral standards on the \citet{Kirkpatrick1991}
system.

\subsubsection{SSO 2.3-m WiFeS IFU}

To measure a preliminary radial velocity, H$\alpha$, and Li~I
$\lambda$6708 line strengths, AP~Col was observed several times during
2011 January - March with the Wide Field Spectrograph (WiFeS) on the
Australian National University 2.3-m at Siding Spring Observatory.
WiFeS \citep{Dopita2007,Dopita2010} is a new dual-beam image slicing
integral field spectrograph that provides a nominal
$25\arcsec\times38\arcsec$ field-of-view with 0.5\arcsec~pixels to two
gratings and camera assemblies simultaneously using a beam splitter,
one `beam' optimized for red spectra, the other for blue.  AP Col was
observed in single-beam mode with the $R7000$ grating, yielding a
resolving power of $R\simeq7000$ and wavelength coverage of
5500--7000\AA.  Following Murphy et al. (2011, in preparation), we
used WiFeS in single-star mode with twice the spatial binning
(1\arcsec~spatial pixels) and optimally extracted and combined the 5
image slices (effectively a 5\arcsec~diameter aperture around the
object) that contain the majority of the stellar flux.

WiFeS has a measured radial velocity precision capability of $\sim$2
km s$^{-1}$ per epoch at this resolution and signal-to-noise
(Table~\ref{tab:optspec}, Murphy et al. 2011, in preparation). We
therefore observed AP~Col seven times on different nights to improve
the mean velocity and check for changes in H$\alpha$ emission line
strength. These observations motivated further high-resolution
observations with the Lick Hamilton Echelle and Keck HIRES Echelle,
discussed in the next sections.

In addition to the $R7000$ observations, AP~Col was observed on 25--26
January 2011 with the $B3000$ and $R3000$ gratings in dual-beam mode,
yielding a resolving power of $R\simeq3000$ and wavelength coverage
from 3400--9560\AA. The spectra from each beam were independently
flux-calibrated and corrected for telluric features, and combined into
a single spectrum.  Details about all the WiFeS observations are
presented in Table~\ref{tab:optspec}.

\subsubsection{Lick Shane 3-m Hamilton Echelle}

Contemporaneous with the low-resolution WiFeS spectra, additional
measurements of radial velocity, H$\alpha$, and Li~I $\lambda$6708 EW
were obtained at the Lick Observatory Shane 3-m telescope with the
Hamilton echelle spectrograph \citep{Vogt1987}, which is located at
the telescope's Coud\'{e} focus. The spectra were bias-subtracted,
flat-fielded, extracted, and wavelength calibrated with ThAr arclamp
spectra.  Further details on data reduction for the Hamilton echelle
with IRAF tasks are outlined in detail in Lick Technical Report
No. 74\footnote{http://astronomy.nmsu.edu/cwc/Software/irafman/manual.html}.

Details about the observations are presented in
Table~\ref{tab:optspec}.  We note that AP~Col, at Decl. -34, is at the
southern limit of what can reasonably be observed from Lick
Observatory; the average airmass during observations was 3.3.

\subsubsection{Keck\,I 10-m HIRES Echelle}

More precise measurements of radial velocity and {\it vsini} of were
obtained on the Keck-I 10-m telescope on Mauna Kea using the HIRES
echelle spectrograph \citep{Vogt1994}.  All HIRES data were reduced
with standard IRAF echelle reduction tasks: data are bias-subtracted,
then flat-fielded with ``wide-decker'' flats (flats taken with twice
the decker height of the science data).  Data are extracted and then
wavelength-calibrated with ThAr arclamp spectra.  Observational
parameters are given in Table \ref{tab:optspec}.  The iodine cell was
in the light path during the observations of AP~Col; as a result,
strong iodine absorption features are present from $\sim$5000-6000\AA.

Archival HIRES observations of the M4.5V star GJ~83.1 were retrieved
from the Keck Observatory
Archive\footnote{http://www2.keck.hawaii.edu/koa/public/koa.php} for
use in radial velocity cross-correlation. GJ~83.1 has a radial
velocity known to be stable to $<$100 m s$^{-1}$ \citep{Nidever2002}.
The archival observations from UT 10 Aug 2009 (PI Haghighipour) were
performed with an identical instrument setup.

\section{Results}

\subsection{Astrometry and Photometry}\label{sec:astrometry}

Based on the $VRI$ photometry from CTIOPI, $JHK_s$ photometry from
2MASS \citep{Cutri2003} (Table \ref{tab:properties}), and the
photometric distance relations from \citet{Henry2004}, we estimate a
photometric distance of 4.6$\pm$0.7 pc.  This is consistent with the
spectrophotometric distance estimates from \citet{Scholz2005} (6.1 pc,
M5.0e) and \citet{Riaz2006} (4 pc, M5 + H$\alpha$).

The astrometric solution for AP~Col was calculated using 14 reference
stars on 158 $V$-band frames taken over 6.48 years, from November 2004
- March 2011.  The resulting absolute trigonometric parallax
(calculated using the pipeline from \citealt{Jao2005}) is
119.21$\pm$0.98 mas (8.39$\pm$0.07 pc), and the relative proper motion
is 342.0$\pm$0.5 mas $yr^{-1}$ at 4.6$\pm$0.1 degrees
(($\mu_{\alpha},\mu_{\delta}$)=(27.33,340.92)$\pm$(0.35) mas
yr$^{-1}$), corresponding to a tangential velocity of 13.60$\pm$0.11
km s$^{-1}$; more details are given in Table \ref{tab:properties}.
The trigonometric distance therefore differs from the photometric
distance estimate by 5-$\sigma$, putting AP~Col $\sim$ 1.5 mag in $V$
above the main sequence. This can be caused by youth and/or unresolved
multiplicity (see \S\ref{sec:multiplicity}).

There is an issue with the particular $V$ filter used between March
2005 and July 2009 (details in \citealt{Subasavage2009} and
\citealt{Riedel2010}) that produces a $\sim$20 mas false astrometric
signal in the R.A. data. An alternative parallax reduction was carried
out using only the data from the other, preferred $V$ filter and was
found to agree with the adopted reduction using all data. We are thus
convinced that our parallax and proper motion are accurate, but our
ability to detect the presence of companions in our astrometric
residuals is thus limited.

\subsection{Spectroscopy}

Based on our CTIO spectrum, AP~Col is an M4.5Ve (accurate to half a
subtype) star with substantial H$\alpha$ emission (Table
\ref{tab:properties}).  This is confirmed with our WiFeS $B/R3000$
spectra, from which we derive a spectral type of M4.5-M5 based on
various spectral indicies and spectrophotometry.

Each of the Hamilton and WiFeS $R7000$ spectra have been
cross-correlated with spectra of stars with known radial velocities to
derive AP~Col's radial velocity and to search for radial velocity
variability between epochs. Each of the Hamilton and WiFeS $R7000$
spectra yield radial velocity measurements for AP~Col with a precision
of roughly 1 km s$^{-1}$ and 2 km~s$^{-1}$, respectively.  Thanks to
the use of the iodine absorption cell, we are able to obtain the best
precision and accuracy on the HIRES radial velocity measurement (see
Table \ref{tab:properties}), and it is most responsible for the
resulting weighted mean radial velocity.

AP~Col's radial velocity appears stable to 1.3 km s$^{-1}$ over 68
days, based on all 9 measurements (Table \ref{tab:optspec}).  Assuming
it is in fact stable, the weighted mean radial velocity is
+22.4$\pm$0.3 km s$^{-1}$.  Our result is consistent with, but more
precise than, the RV reported by \citet{Scholz2005},
(V$_{rad}$=+67$\pm$30 km s$^{-1}$). This is not surprising considering
the low resolution ($R\simeq600$) of the spectra they used.

AP~Col is a known flare star, as reported in \citet{Ball1995}, and we
detected strong H$\alpha$ and numerous other emission lines in its
optical spectra (listed in Table \ref{tab:emlines}). Indeed, the Lick
and WiFeS $R3000$ data show AP~Col in the midst of an energetic
outburst (H$\alpha$ EW $\approx$35\AA). This outburst will be
discussed in detail in a future publication (Melis et al. 2011,
in preparation).

\subsection{Multiplicity}\label{sec:multiplicity}

As determined in \S\ref{sec:astrometry}, AP~Col lies $\sim$1.5
magnitudes above the main sequence for a star of spectral type M4.5.
Overluminosity can be attributed to at least one of three things: 1)
multiplicity (where the brightness of the star is actually the
combination of two or more stars); 2) youth (where the star is still
gravitationally contracting onto the main sequence), or 3) high
metallicity (where the star is not brighter, but redder than a typical
main sequence star of the same mass and luminosity, much as subdwarfs
are bluer than corresponding main-sequence stars).

Given that AP~Col lies well above even the high-metallicity envelope
of stars within 10 pc, we can reasonably discard the last option.
Multiplicity can conspire to make a system appear up to 41\% closer as
an equal-luminosity binary, or 73\% closer as an equal-luminosity
trinary. The measured discrepancy, 81\%, cannot thus be explained by
an equal luminosity binary, even when the full 2-$\sigma$ systematic
error on the photometric distance -- 30\%, \citealt{Henry2004} -- is
assumed.  If the multiple system is close enough, the stars can
tidally interact and force synchronous rotation, which can maintain
fast rotational velocities and the resulting chromospheric activity
until the system is far older than what is considered `young'.

To address the potential multiplicity of AP~Col, we work our way
inwards.  Wide surveys of AP~Col on SuperCOSMOS plate scans
\citep{Hambly2001} reveal no wide companions to the star within
15\arcmin\ and brighter than $SERC(J) = 21$,$SERC(I)=19$; within that
range only the star 2MASS J06052273-3429245 has noticeable proper
motion, (-62.38,-119.63)$\pm$(7.02,9.95) mas yr$^{-1}$, but this
proper motion is not similar to AP~Col.  2MASS $JHK_S$ images show no
genuine companions down to $\sim$3\arcsec; the apparent close
companions are actually `glints', detector artifacts from internal
reflections in the
camera\footnote{http://www.ipac.caltech.edu/2mass/releases/allsky/doc/sec4\_7.html}.

\begin{figure}
\center
\includegraphics[angle=0,scale=.6]{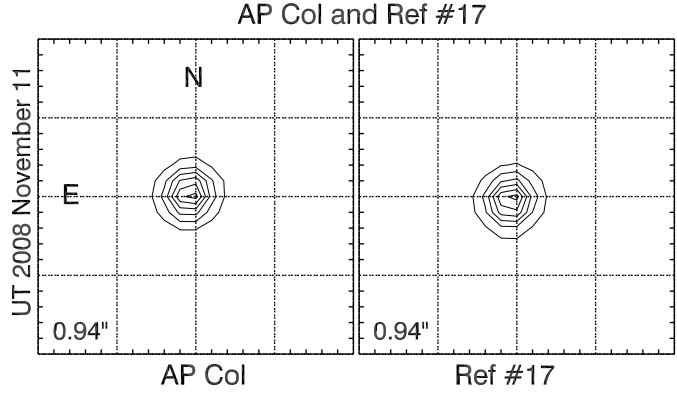}
\caption{The field immediately surrounding AP~Col and the closest
  reference star, in $V$ from the CTIO 0.9m, on a night with good
  seeing (FWHM=0.94\arcsec).  There is no sign of a companion to
  AP~Col down to $\Delta$V=5 mag, or any visible difference between
  the PSF of AP~Col and Ref \#17.
  \label{fig:contours}}
\end{figure}

The highest resolution seeing-limited image of AP~Col from the CTIOPI
frames (FWHM 0.94\arcsec, taken on UT 11 NOV 2008) shows no signs of
companions to AP~Col down to separations of 0.9\arcsec and $\Delta
V$=5, nor any sign of elongation from an unresolved companion; other
images in all three filters show nothing to approximately that limit
as well\footnote{CTIOPI exposes target stars to roughly the same ADU
limit regardless of filter; the limits in $\Delta V$, $\Delta R$ and
$\Delta I$ are thus roughly identical.} (Figure \ref{fig:contours}).

\citet{Bergfors2010} lucky-imaged 124 M dwarfs from the
\citet{Riaz2006} sample, including AP~Col.  Their images, taken in
early November 2008, detected no companions to AP~Col at angular
separations between 0.1\arcsec-6.0\arcsec, and a magnitude difference
of $\Delta z' \leq$ 2~mag at the smallest separations.  Independently,
Mason \& Hartkopf (2011, private communication) observed AP~Col in
early March 2006 with the USNO Specklecam on the CTIO 4-m and also
obtained a null result, with $\Delta vis \leq$~3 at a separation of
0.05\arcsec--1.0\arcsec.

Further constraints using our CTIO astrometric data are problematic
due to the issue with the $V$ filters mentioned in
\S\ref{sec:astrometry}. While we see no evidence for a companion in
the astrometric residuals, we can only constrain the possible
multiplicity of AP Col to objects that would produce a photocentric
shift smaller than 20 mas in right ascension or 6 mas in
declination\footnote{Because we are measuring the motion of the
photocenter, we are insensitive to equal-mass, equal-luminosity
binaries whose photocenter would not shift at all, but we {\it can}
rule out unequal companions.}.

\begin{figure}
\center
\includegraphics[angle=0,scale=.6]{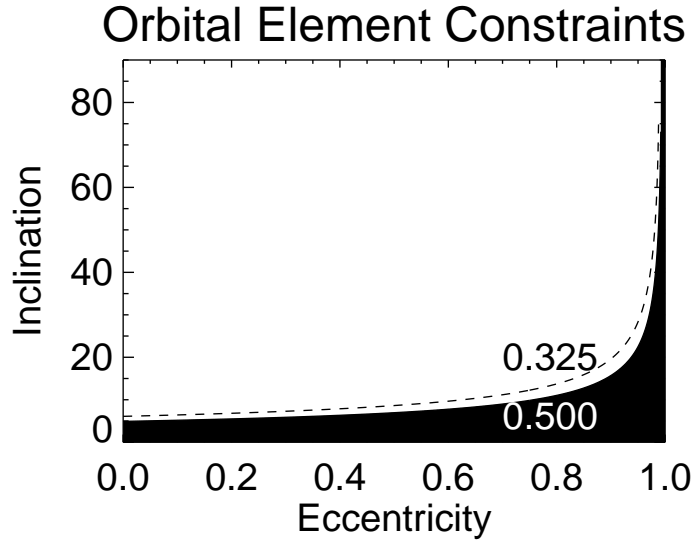}
\caption{Companion detection limits, given the null results of our
  companion search down to 0.05\arcsec~separations, radial velocities
  stable to 1.3 km s$^{-1}$, and assuming a maximum separation of 0.42
  AU corresponding to 0.05\arcsec~at 8.4 pc. Below the dashed line, a
  0.075\msun~brown dwarf ($M_{tot}$ = 0.325\msun) could be hidden;
  within the filled region, two M4.5V stars ($M_{tot}$ = 0.5\msun, the
  most likely scenario to explain the overluminosity) could be hidden.
  Such eccentricities and inclinations are unlikely, and we conclude
  that AP~Col is a single star.
  \label{fig:threshold}}
\end{figure}

These visual limits, particularly the lucky and Speckle Imaging, set
strict limits on the size of a companion's orbit.  \citet{Henry1999}
suggests an M4.5V main sequence star has a mass of roughly 0.25 \msun;
to best explain the overluminosity requires a twin 0.25
\msun~star\footnote{As mentioned earlier, additional components would
better provide the additional flux, but would necessitate a higher
system mass and larger velocity variations}.  With those masses, the
longest period circular orbit that could be hidden within
0.05\arcsec~(0.42 AU) is 0.38 years, for which the velocity amplitude
would be 33 km s$^{-1}$.  We have already established that the radial
velocity of AP~Col is stable to within 1.3 km s$^{-1}$ over the 9
epochs and 68 days of RV observations ($\sim 1/2$ of the maximum
orbital period) with cadences as short as 1 day; this significantly
limits the inclinations and eccentricities in which a companion could
remain undetected, as shown in Figure \ref{fig:threshold}.  Low
inclinations can particularly be ruled out given that we have measured
the {\it vsini} rotational velocity as 11 $\pm$ 1 km s$^{-1}$, which
would translate to nearly 300 km s$^{-1}$ (near break-up speed) if the
star's rotational axis were sufficiently inclined to hide a
low-eccentricity orbit.  We are thus convinced that AP~Col has no
stellar-mass companions; substellar companions may still be present
but cannot explain AP~Col's elevation above the main sequence.  We
conclude that the activity and overluminosity we see in AP~Col are
intrinsic to the star itself (and that it is {\it not} necessarily a
main sequence star with a mass of 0.25~\msun), and not the result of a
stellar companion.

\subsection{Youth}

\subsubsection{Isochronal age}

Stars that have not yet reached the main sequence (i.e., equilibrium
between gravitational collapse and thermonuclear fusion) will appear
overluminous compared to other stars of the same color because their
photospheric surfaces are larger than main-sequence stars.  Such young
stars are therefore elevated relative to the main sequence on an HR
diagram.

To properly compare AP~Col to other known young stars without relying
on theoretical models requires accurate data.  Therefore, we compiled
a catalog of known young stars in associations from
\citet{Zuckerman2004} and \citet{Torres2008}.  Position, proper motion
and photometry for each star were extracted from the All Sky Compiled
Catalog (ASCC, \citealt{Kharchenko2001}), while trigonometric
parallaxes\footnote{Unlike the generally kinematic distances in
\citet{Zuckerman2004} and \citet{Torres2008}, trigonometric parallaxes
are insensitive to any accidental mis-identifications with
associations.  While adding kinematic distances to high-confidence
members {\it would} improve the number of points used in the fits,
there are very few known young M dwarfs in these associations.} were
obtained from multiple sources, principally {\it Hipparcos}
\citep{van-Leeuwen2007}, the General Catalog of Trigonometric
Parallaxes \citep{van-Altena1995}, \citet{Weinberger2011}, and
unpublished data from the CTIOPI observing program.  To produce
empirical isochrones for nearby associations, multiples, suspected
multiples, and obvious outliers (generally noted as such in the
Hipparcos catalog, \citet{Zuckerman2004}, or \citet{Torres2008}) were
removed from the subset of young stars with parallaxes.  The sample
was then approximated by fifth-order polynomials to form rough
isochrones, which are listed in Table \ref{tab:polynomials}.

\begin{figure}
\center
\includegraphics[angle=90,scale=.7]{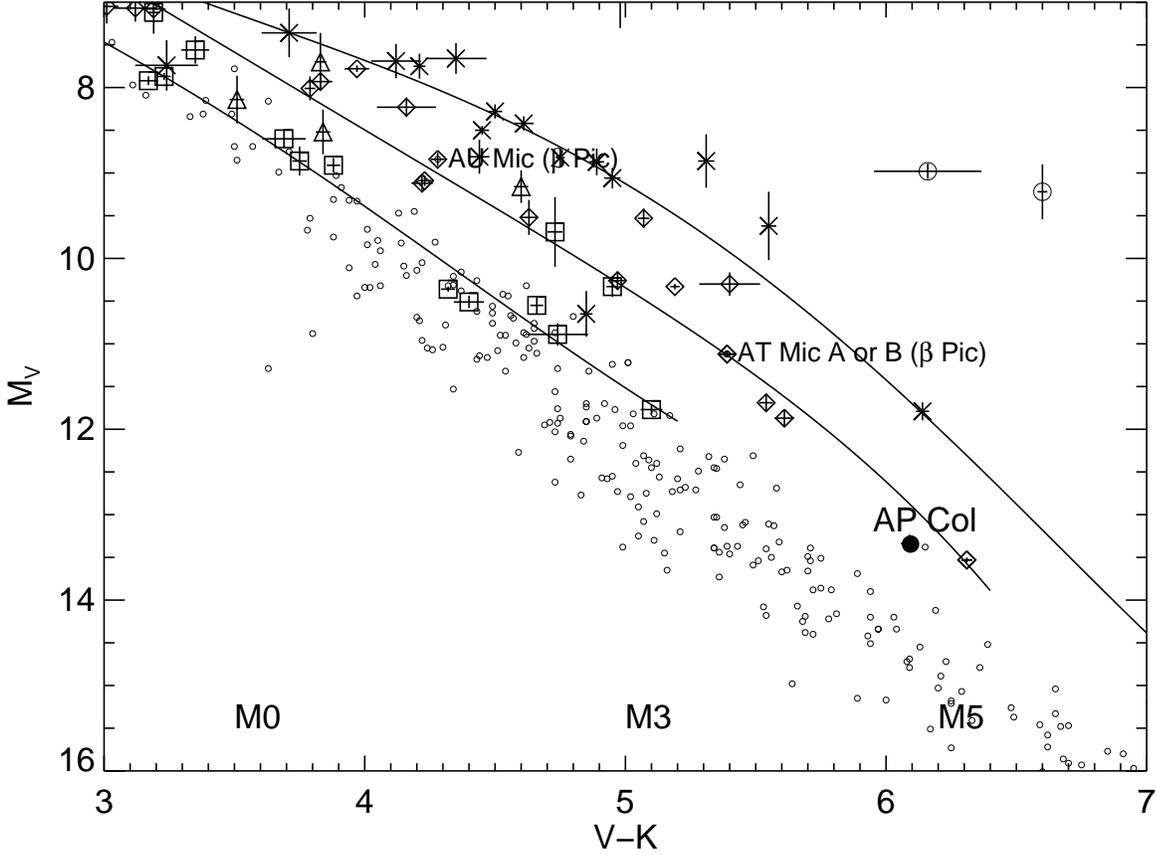}
\caption{AP~Col (filled circle) plotted relative to the RECONS 10 pc
  sample (the elevated 10 pc object near AP~Col is EQ Peg B, see the
  text), with error bars smaller than the plotted symbol.  Also
  plotted are members of nearby young associations from
  \citet{Zuckerman2004} and \citet{Torres2008}: $\epsilon$~Cha (large
  open circles), TW~Hya (Xs), $\beta$~Pic (diamonds), Tuc-Hor
  (triangles), and AB~Dor (squares).  Fifth order fits are plotted for
  (top to bottom) TW~Hya, $\beta$ Pic, and AB~Dor.  No attempt has
  been made to split any unresolved binaries among the associations
  other than the AT Mic A\&B and TWA 22 A\&B systems, which both
  provide overlapping points.  AP~Col appears to be older than (but
  consistent with), $\beta$~Pic; and younger than (but consistent
  with) AB~Dor, although at such red colors and low temperatures, none
  of the association memberships or isochrone fits are well defined.
  \label{fig:isochrones}}
\end{figure}

Figure \ref{fig:isochrones} shows AP~Col and the other members of the
RECONS 10 pc sample (stars in systems with parallaxes greater than 100
mas and errors less than 10 mas, \citealt{Henry2006}).  Known young
stars with trigonometric parallaxes are also shown on the graph, along
with error bars based on their parallactic and photometric errors
(AP~Col's errors are smaller than the plotted symbol), and polynomial
fits to three of the associations.  As the fifth-order polynomials are
highly sensitive to the colors of their most extreme members, we have
not plotted them past the reddest young star in each association.  The
$\epsilon$ Cha isochrone relies on too few points to be reliable, and
there are no quality Tuc-Hor members cooler than M1 (the single point
near AU Mic appears to be a binary).  The reddest object in
$\beta$~Pic isochrone is TWA 22AB, originally misclassified as a TW
Hya member, with a parallax reported by \citet{Teixeira2009}. Only the
TW~Hya association extends redder than the plot in Figure
\ref{fig:isochrones}; its reddest member is TWA 27 ($V-K_S$=8.25,
M$_V$=16.60), with parallaxes in \citet{Gizis2007}, \citet{Biller2007}
and \citet{Ducourant2008}.

As can be seen in Figure \ref{fig:isochrones}, AP~Col is clearly older
than members of the $\epsilon$ Cha cluster ($\sim$6 Myr), and
consistent with but likely older than $\beta$~Pic ($\sim$12 Myr).
There are no comparably red AB Dor or Tuc-Hor members with known
parallaxes; despite this, the position of AB~Dor at the
high-metallicity upper envelope of the main sequence suggests that
AP~Col is younger than the $\sim$70 Myr AB~Dor association.

\subsubsection{Lithium}

The Li {\sc I} $\lambda$6708 equivalent width of AP~Col is
0.28$\pm$0.02\AA, the weighted mean of all our high resolution
spectral measurements.  Lithium is easily destroyed but not readily
produced by stellar thermonuclear fusion, and is thus only detected in
the photospheres of objects that have not yet consumed their
primordial supply of the element. This is usually interpreted as a
consequence of youth, or in the case of brown dwarfs, because their
cores never reach the temperatures necessary to fuse it. Lithium
depletes fastest in mid-M stars, where it is thought that the
persistence of full convection throughout the star's evolution to the
main sequence means that all the lithium is cycled through the core
and quickly destroyed once temperatures rise high enough
\citep{Jeffries2001}.  Larger, hotter stars develop radiative cores
that take longer to deplete their lithium and can trap it in their
photospheres for a billion years; cooler fully-convective stars like
AP~Col have longer nuclear burning timescales.  Thus, as a coeval
stellar population ages, the temperature range of lithium-depleted
stars widens around the mid-M stars, with a particularly sharp drop on
the cooler side.  The detection of lithium in a K or M-type star is
therefore a strong indicator of youth.

\begin{figure}[p] 
\center
\includegraphics[width=\textwidth]{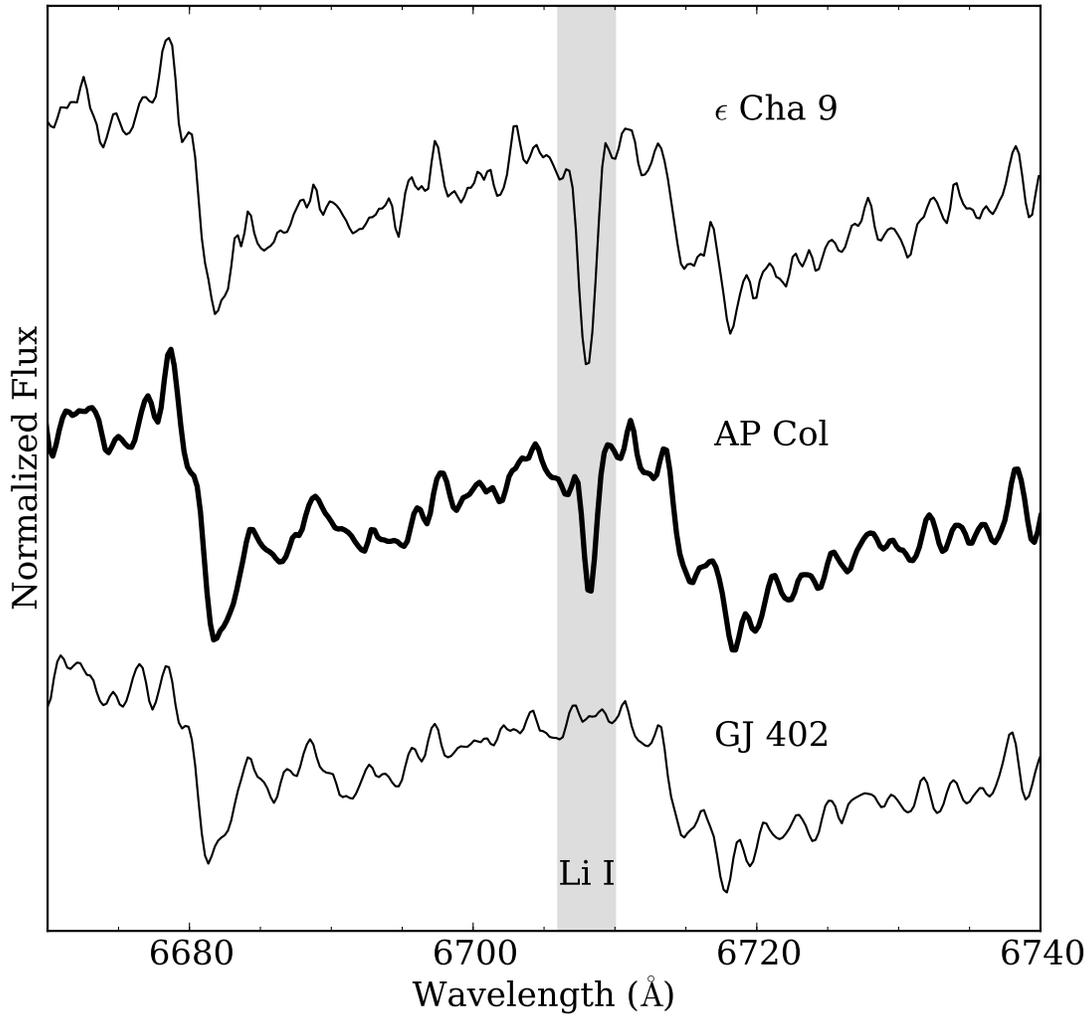} 
\caption{WiFeS $R7000$ spectrum of AP~Col showing strong lithium
  absorption (Li I $\lambda$6708). The strength of this absorption is
  intermediate between that of the young ($\sim$6 Myr old)
  $\epsilon$~Cha member $\epsilon$Cha 9 and the older field dwarf GJ 402.
  \label{fig:lithium}}
\end{figure}

By way of comparison, Figure \ref{fig:lithium} presents WiFeS spectra
in the vicinity of the Li {\sc I} $\lambda$6708\AA~feature for AP~Col,
$\epsilon$~Cha 9 (a $\sim$6 Myr old member of the $\epsilon$~Cha
association), and GJ 402 (a field radial velocity standard), all of
the same approximate spectral type.  The strength of the lithium
absorption in AP~Col is intermediate between the young $\epsilon$ Cha
member and the old field dwarf, and suggests an intermediate age.

The dependence of lithium burning efficiency on stellar age and
temperature leads to the concept of a lithium depletion boundary
(LDB), where only a small change in stellar mass and temperature can
lead to the appearance or disappearance of the Li {\sc I}
$\lambda$6708 feature \citep{Song2002}.  For instance, at the age of
the $\beta$~Pic association ($\sim$12 Myr), the cool edge of the LDB
lies at the spectral type M4.5 \citep{Song2002,Torres2003}, while at
the age of the Pleiades ($\sim$100-130 Myr) it has shifted to M6.5
\citep{Stauffer1998,Barrado-y-Navascues2004}. While \citet{Yee2010},
\citet{Song2002} and others have noted that lithium depletion ages are
systematically larger than those derived by isochrone fitting or
kinematic expansion ages, relative age ranks are not affected.

We plot in Figure~\ref{fig:lithium_groups} the lithium measurements
for AP~Col, several of the young stellar associations in the solar
vicinity from \citet{da-Silva2009}, and the young cluster IC~2391 from
\citet{Barrado-y-Navascues2004}. The $\beta$~Pic LDB at M4.5 (dashed
line) is clearly visible as the discontinuity in equivalent width
around 3300~K, while the IC~2391 lithium depletion boundary at
$\sim$M5 (3200K) is shown as a dotted line.  AP~Col lies between the
two boundaries at $T_{\rm eff}\simeq3250$ K (based on $V-I$ color and
the conversion of \citet{Kenyon1995}), implying an age between
$\beta$~Pic ($\sim$ 12 Myr) and IC~2391 \citep[$50 \pm 5$~Myr lithium
depletion age,][]{Barrado-y-Navascues2004}, though consistent with
either.

\subsubsection{Low-gravity features}

\begin{figure}
\centering
\includegraphics[width=0.7\textwidth]{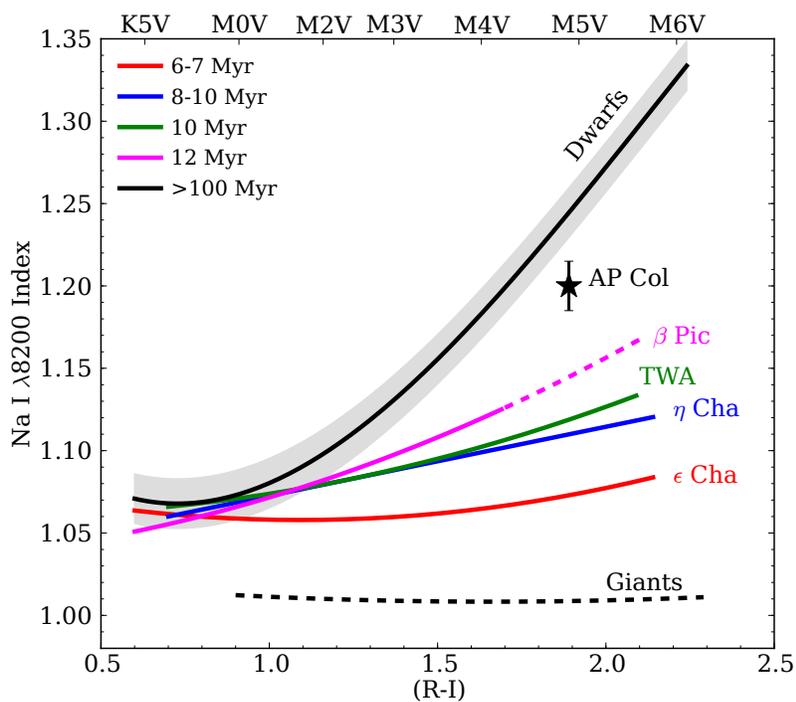}
\caption{Na~I index trends for various young, local associations from
  \citet{Lawson2009b}. The shaded band represents the variation around
  the mean dwarf trend seen in \citet{Lyo2004a}. The value for AP~Col
  has been derived from our WiFeS R3000 spectra, smoothed to the
  approximate resolution of the \citeauthor{Lawson2009b} data. The
  error bar shows the variation observed between the four exposures.
  AP~Col has an intermediate gravity, suggestive of an age between
  12--100 Myr.}
   \label{fig:gravgroups}
\end{figure}

Certain spectral features are sensitive to the surface gravity of a
star, and may therefore be used as age proxies for stars still
contracting toward the main sequence \citep{Hayashi1966}, at least for
relative dating \citep[e.g.][]{Lawson2009b}. The Na~I
$\lambda$8183/8195~doublet is particularly useful for comparative
gravity studies \citep[e.g.][]{Lawson2009b,Murphy2010}. For spectral
types cooler than $\sim$M3, there is a marked decrease in the strength
of the Na~I doublet between dwarfs (strong), pre-main-sequence stars
(intermediate), and giants (weak/absent).  To measure the strength of
Na I absorption, we adopt the spectral index of \citet{Lyo2004a},
formed by the ratio of the average flux in two 24 \AA wide bands: one
on the doublet and one on the immediately adjacent pseudo-continuum.
Figure \ref{fig:gravgroups} shows the Na~I $\lambda$8200~index for AP
Col compared to the mean trends of other young associations in
\citet{Lawson2009b}.  To match the resolution used in that study, we
have smoothed and resampled the WiFeS $R3000$ data to $R\sim900$ and
the same wavelength scale used by \citet{Lawson2009b}.  Although close
to the dwarf locus, AP~Col nevertheless lies at intermediate gravities
between $\beta$~Pic and field dwarfs; we can again constrain the age
to greater than that of $\beta$~Pic and less than the Pleiades, whose
M-type members have gravity features indistinguishable from field
stars \citep{Slesnick2006}.

Alkali metal lines such as Na~I can also be affected by stellar
activity, where emission fills in the absorption line cores, leading
to lower EWs \citep{Reid1999}.  Our WiFeS observations span a factor
of three in H$\alpha$ EW, but no correlation between that activity
indicator and our Na~I doublet EW measurements could be found.
\citet{Slesnick2006b} notes that the Na~I $\lambda$8183/8195~doublet
can be affected by telluric absorption over the region 8161--8282~\AA,
leading to artificially low Na~I index values for stars observed at
large airmasses.  Our WiFeS $R3000$ spectra were observed at $\rm
sec(z)\simeq1$; nevertheless we have checked the telluric correction
of the spectra and find no excess that could affect the index
measurements.

\subsubsection{v$sin$i}

Young stars are expected to rotate rapidly, with decreasing rotation
as they age.  \citet{Reiners2009} suggest {\it vsini} $>$ 20 km
s$^{-1}$ is a rapidly rotating M star.  From the Keck-I HIRES spectra,
we measure a {\it vsini} = 11 $\pm$ 1 km s$^{-1}$, indicating that AP~Col
is not necessarily a rapidly rotating star.  Because it is not in a
binary system, this spin is not due to tidal synchronization with a
companion; it is a remnant of the star's formation.

While gyrochronology relations exist \citep[e.g.][]{Mamajek2009} for
solar-type stars, none have been developed for M dwarfs, nor stars
with saturated X-ray emission such as AP Col.  Gyrochronology cannot
(yet) be used to estimate an age for AP~Col.

\subsubsection{Activity: X-ray emission, H$\alpha$ emission, flares, and photometric variability}

Young stars are known to have large amounts of X-ray emission,
H$\alpha$ emission, flares, and  photometric variability, when compared
to field stars.  All of these indicate chromospheric activity, fast
rotation rates and powerful magnetic fields.  This activity also
manifests itself photometrically, and young stars have long been
recognized for long-term large-amplitude variations and flares since
they were first discovered as ``Orion-type'' variables.

Unfortunately, M dwarfs undergo a long adolescence, and as recognized
by \citet{Zuckerman2004} and \citet{Hawley1996}, X-rays remain
saturated at $log$(L$_x$/L$_{bol})\sim-3$ in M dwarfs past the age of
the Hyades ($\sim$600 Myr). H$\alpha$ persists at varying levels for
equally long times, and the variability and flaring continues long
afterwards into the intermediate-age UV Ceti flare stars.  Thus, these
activity indicators are necessary but insufficient indicators of
youth, fairly irrelevant in the face of the measured lithium
absorption discussed above, and are discussed mainly for completeness.

AP~Col is cross-identified as the ROSAT All Sky Survey object 1RXS
J060452.1-343331, and has $log$(L$_x$/L$_{bol}$)=$-$2.95$\pm$0.16 and
$log$(L$_x$)=28.49 (17\% error).  These values match \citet{Riaz2006}
and agree with the range of X-ray variability published by
\citet{Scholz2005}, $log$(L$_x$/L$_{bol}$)=$-$3 to $-$4.

As seen in Table \ref{tab:optspec}, during our spectroscopic
observations the H$\alpha$ EW of AP~Col varied from $-$6\AA~in apparent
quiescence, to $-$35\AA~during the strong flare on 2011 Jan 25, with an
average EW of $-$9.1$\pm$5.2\AA, in agreement with the $-$12.1\AA~single
epoch measurement published by \citet{Riaz2006}.

AP~Col is a known UV Ceti flare star, and \citet{Ball1995} observed 5
flares over 9 hours of $U$-band observations, the largest of which was
2.5 magnitudes above background.  A 12-hour X-ray flare was also
measured by ROSAT and presented by \citet{Scholz2005}; during this
event AP~Col increased in X-ray luminosity by roughly an order of
magnitude and slowly dropped back to normal levels.  We have also
measured an energetic white-light flare in Lick Hamilton Echelle data
taken 25 January 2011 (Melis et al.\ 2011, in preparation).

\begin{figure}
\center
\includegraphics[angle=90,scale=.33]{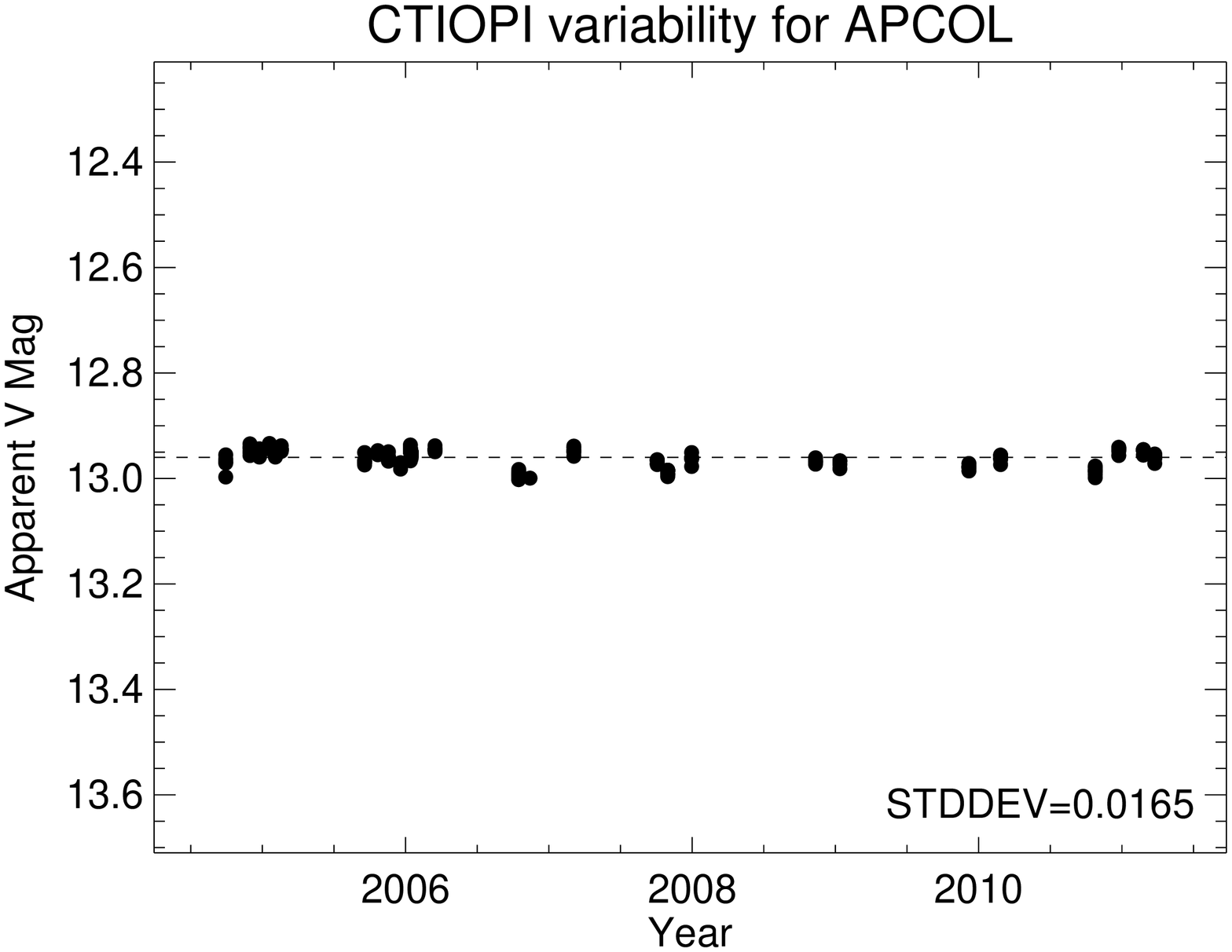}
\includegraphics[angle=90,scale=.33]{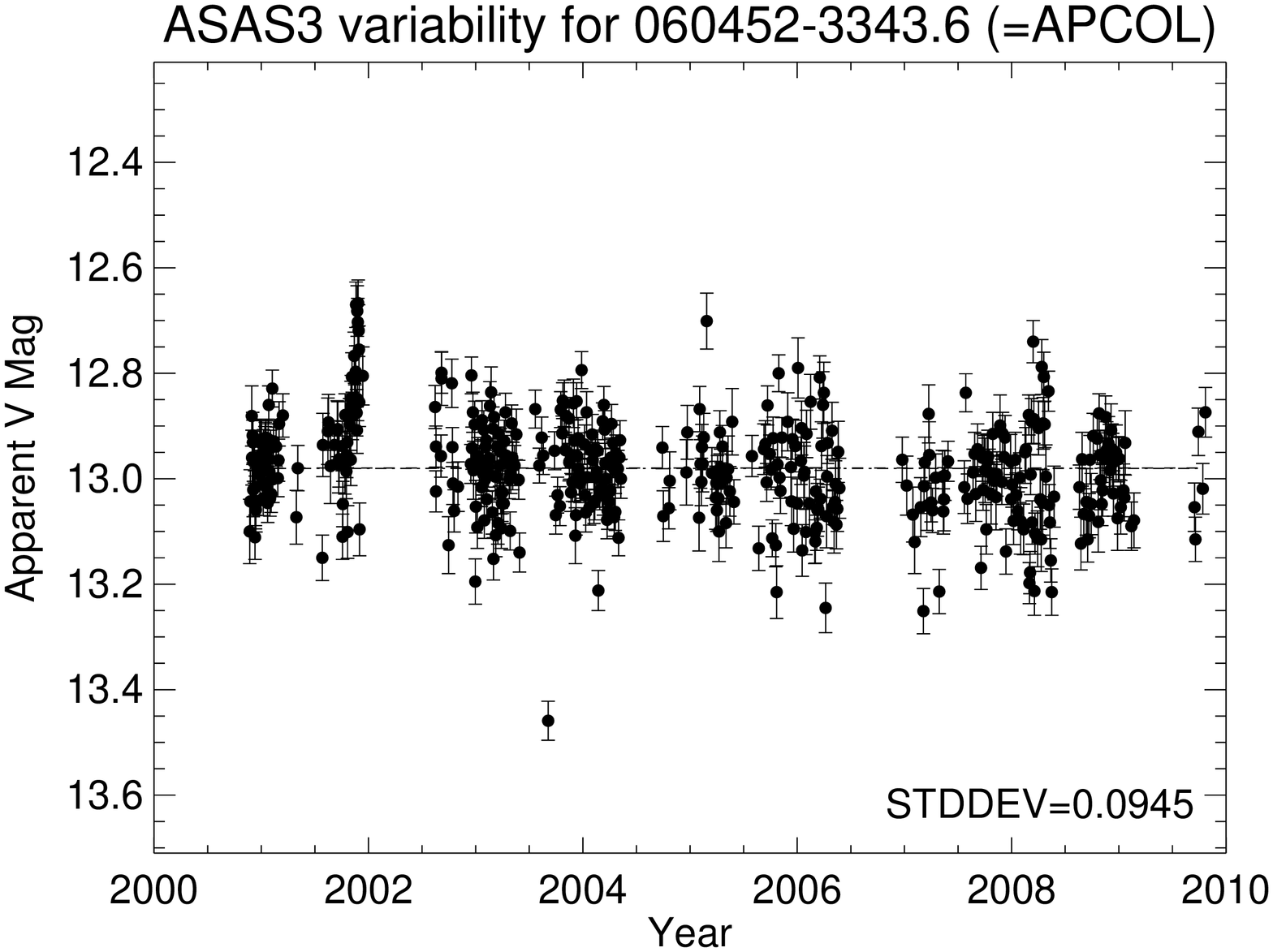}
\caption{AP~Col relative photometric variability in the $V$ filter
  observations from CTIOPI (left) and ASAS (right).  ASAS has lower
  photometric accuracy than CTIOPI relative photometry.  Neither time
  series shows variability or flares.
  \label{fig:variability}}
\end{figure}

\begin{figure}
\center
\includegraphics[angle=0,scale=.6]{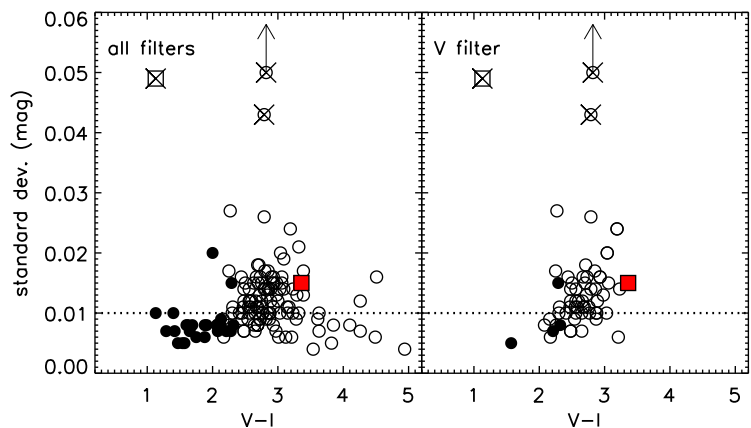}
\caption{AP~Col (filled square) relative variability as compared to
  other stars studied during CTIOPI \citep{Jao2011}.  Apart from being
  somewhat redder than other stars observed in the $V$ filter, AP~Col
  is unremarkable.
   \label{fig:jaovar}}
\end{figure}

Finally, we have measured the relative photometric variability of AP
Col using the $V$ filter data from the CTIOPI astrometric frames.  The
standard deviation of the variability is 1.7\%, (Figure
\ref{fig:variability}).  As shown in Figure \ref{fig:jaovar}, this is
typical when compared to field M dwarfs observed during CTIOPI
\citep{Jao2011}.  The 158 CTIOPI astrometry frames constitute a total
observing time of 14715 seconds over 27 nights (4 hours, with a median
time of 450 seconds in five observations per night) spanning 6.48
years.  We additionally checked the ASAS3 database
\citep{Pojmanski1997} for photometry on AP~Col, and find no evidence
of large flares in their dataset (see Figure \ref{fig:variability})
either.  We report our own variability in Table \ref{tab:properties},
as our 0.91-m telescope aperture lends itself to better photometry
than the 8-cm ASAS telescopes.

\subsubsection{IR detection}

Young stars often have protostellar disks that show up as near- and
mid-IR excesses.  These tend to vanish within $\sim$10 Myr
\citep{Haisch2001,Haisch2005}.  IRAS and WISE photometry from the
preliminary data release show no obvious signs of infrared excess
around AP~Col.  This is not unexpected if AP~Col is older than
$\sim$10 Myr, as suggested by our other age indicators.

\section{Conclusions}

The balance of the age indicators place AP~Col somewhere between the
ages of $\beta$~Pic and IC~2391, or $\sim$12 Myr to $\sim$50 Myr
following \citet{Torres2008} and \citet{Barrado-y-Navascues2004}. To
give the discovery of AP~Col context, we compare it to the other 255
stellar systems known within 10 pc as of January 1, 2011 \citep[and
updates at {\it www.recons.org}]{Henry2006}.  As shown in the
color-magnitude diagram of Figure \ref{fig:isochrones}, AP~Col is one
of only a few red dwarfs noticeably elevated above the main sequence,
with a location of $M_V$ = 13.34, $V-K$ = 6.09.

Three of the elevated points within 10 pc are in one system, comprised
of \object{AU Mic}, \object{AT Mic A}, and \object{AT Mic B} (9.9 pc;
note that the AT Mic A+B point is actually the overlap of two similar
points --- the two stars have virtually identical $V$ and $K$
magnitudes).  This triple is one of the prototypical members of the
$\beta$~Pic association \citep{Barrado-y-Navascues1999} with an age of
$\sim$12 Myr, and is remarkable as the youngest of the 256 systems
known within 10 pc.

The elevated point near AP~Col in Figure \ref{fig:isochrones}
represents \object{GJ 896 B} (6.3 pc), otherwise known as EQ Peg B, at
$M_V$ = 13.38, $V-K$ = 6.15.  Both \object[GJ 896 A]{EQ Peg A} and B
are known to flare, have H$\alpha$ in emission, and emit X-rays
\citep{Robrade2004}.  Both components have also been reported to have
companions \citep{Delfosse1999}, but a private communication from the
first author of that study indicated that neither spectroscopic
companion was confirmed.  Thus, we are left with a mystery: the EQ Peg
system exhibits some indicators of youth and neither component is
known to be multiple, but the A component at $M_V$ = 11.24, $V-K$ =
4.95, is not significantly elevated above the main sequence, while the
B component is.

Another potentially young star within 10 pc is \object{GJ 393} (7.1
pc), which \citet{Torres2008} report as a member of AB Dor.  However,
the star has weaker X-ray and NUV emission \citep{Rodriguez2011}, no
H$\alpha$ emission, and slow rotation ({\it vsini} $<$ 3,
\citealt{Rodriguez2011}), all more typical of older stars.  We suspect
it is a high metallicity field star with space velocities
coincidentally similar to AB~Dor, and note that the AB~Dor isochrone
happens to lie along the high-metallicity envelope of the main
sequence (Figure \ref{fig:isochrones}). \citet{Lopez-Santiago2009}
also report GJ 393 as an interloping main sequence star.

Ultimately, the statuses of GJ 393 and GJ 896 AB are still uncertain.
AP~Col is now the second youngest system within 10 pc, and the closest
one, at 8.4 pc from the Sun.

\subsection{Argus / IC~2391 membership}

\begin{figure}
\center
\includegraphics[width=\textwidth]{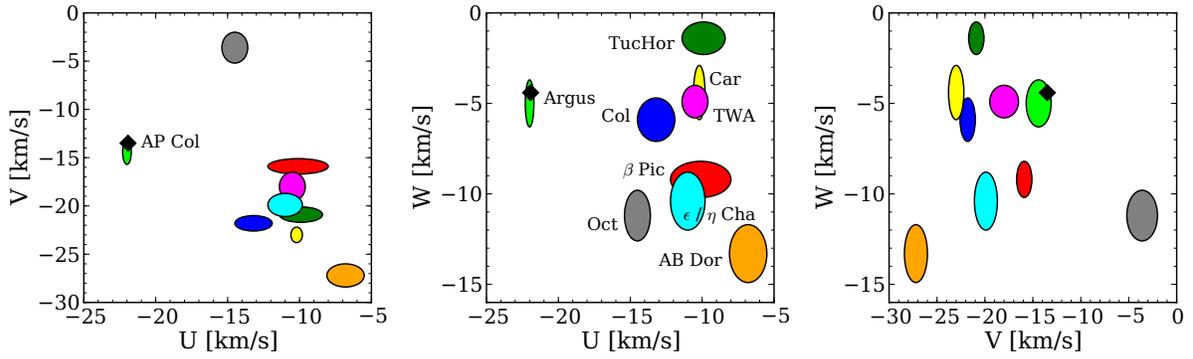}
\caption{AP~Col (diamond) plotted in {\it UVW} phase space relative to
  the associations in \citet{Torres2008}.  Ellipses show the velocity
  dispersions of the various groups.  Error bars (within the diamond)
  reflect the accuracy of our measurements.  The space motion of AP
  Col is consistent only with the Argus association.
  \label{fig:UVW}}
\end{figure}

\begin{figure}
\center
\includegraphics[width=\textwidth]{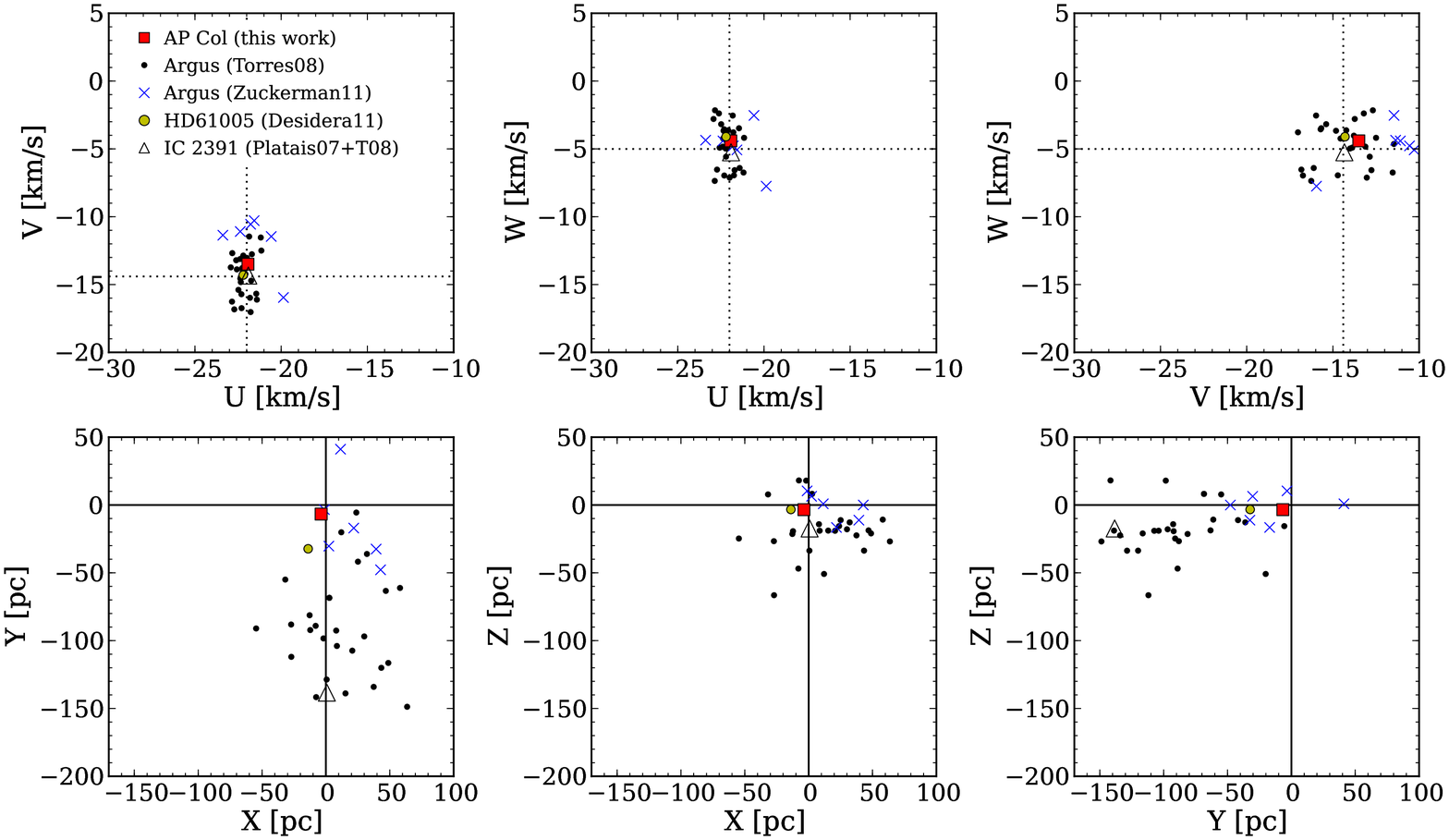}
\caption{AP~Col (filled square) plotted in {\it UVWXYZ} phase space
  relative to the Argus Association as defined by \citet{Torres2008}
  (small points). Recent new members found by \citet{Desidera2011} (HD
  61005, large circle) and \citet{Zuckerman2011} (crosses) are also
  plotted. As shown by \citet{Torres2008}, the young open cluster
  IC~2391 (open triangle, data from \citet{Platais2007,Torres2008})
  appears to be kinematically and spatially associated with Argus as
  well as having a similar age (see text).
  \label{fig:kinematics}}
\end{figure}

One additional point of great interest is that the kinematics of AP
Col are an excellent match for the $\sim$40 Myr old Argus association
defined in \citet{Torres2003} and updated in \citet{Torres2008}.
Combining the radial velocity with the CTIOPI parallax and proper
motion data allows us to calculate the $UVWXYZ$ phase-space positions
for AP~Col.  These are $(X,Y,Z)=(-3.72, -6.70, -3.41) \pm (0.04, 0.08,
0.04)$ pc and $(U,V,W)=(-21.98, -13.58, -4.45) \pm (0.17, 0.24, 0.13)$
km~s$^{-1}$.  This places AP~Col only 1.0~km s$^{-1}$ from the mean
velocity of the Argus Association, $(U,V,W) = (-22.0, -14.4, -5.0) \pm
(0.3, 1.3, 1.3)$ km~s$^{-1}$ \citep{Torres2008}. As seen in Figure
\ref{fig:UVW}, Argus is the only possible match for AP~Col among the
nine nearest known associations, given its observed kinematics. In
Figure~\ref{fig:kinematics} we plot the phase space location of AP~Col
relative to other proposed Argus members from \citet{Torres2008},
\citet{Desidera2011}, and \citet{Zuckerman2011}. In addition to
congruent kinematics, AP~Col occupies a region of $XYZ$ space on the
outskirts of known members. Argus is likely much larger than the
volume traced by known members, and new members continue to be
identified. At an age of $\sim$40~Myr, Argus also has an isochronal
age in the middle of the range expected from our gamut of age
indicators.

Although suggestive, kinematics alone are insufficient to argue
membership. Unfortunately, there are no previously known M-type Argus
members to which we can directly compare AP~Col. However, the apparent
link between Argus and the young open cluster IC~2391 can provide some
insight and reduce the range of possible ages for AP~Col.

IC~2391 has similar kinematics (including its ``special U velocity'',
\citealt{Torres2008}), spatial location\footnote{IC~2391's location is
shown as derived from thirteen \citet{Torres2008} members \citep[drawn
from the list of][]{Platais2007} with Hipparcos astrometry.  The
\citet{Torres2008} distance, 139.5~pc, gives the best kinematic
agreement between Argus and IC 2391 but differs from the
\citep{Platais2007} best-fit main sequence distance of 156~pc. The
mean Hipparcos distance is between those two values, at 146~pc.}
(Figure~\ref{fig:kinematics}), and a similar age to Argus
\citep{Torres2008,Makarov2000}.  As such, the field members of Argus
may be `evaporated' members of IC~2391 stripped free by internal and
external interactions with other stars, or distant products of the
same filament of gas that eventually became IC~2391.  Argus is
projected to extend over a huge volume of space, reaching from the
center of IC~2391 (distance $\sim139\pm7$ pc, \citealt{Torres2008}) to
as close as 11 pc \citep{Zuckerman2011}, and, with the discovery and
characterization of AP~Col, perhaps even 8.4 pc.

\begin{figure}
\center
\includegraphics[width=\textwidth]{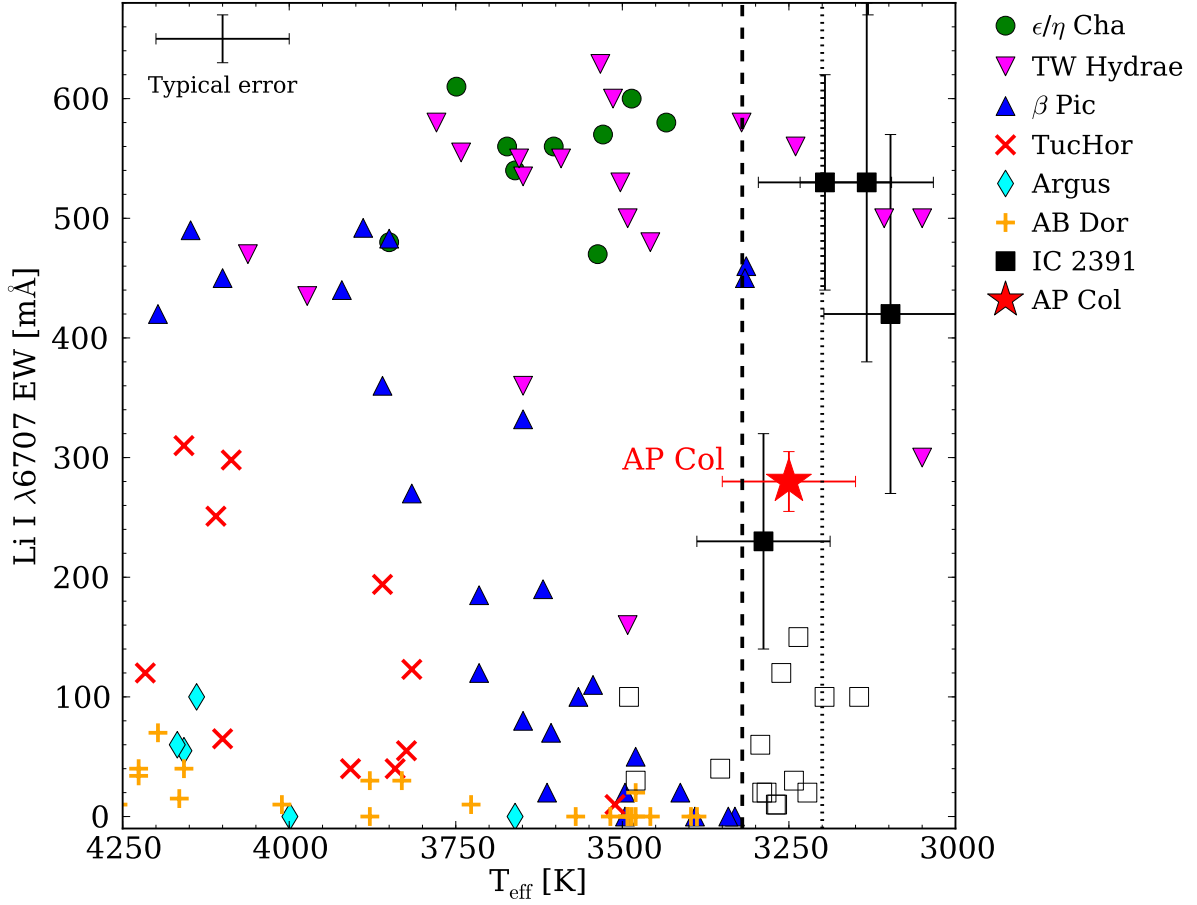}
\caption{Li {\sc I} $\lambda$6708 absorption for AP~Col (filled star)
  relative to the young local association members measured by
  \citet{da-Silva2009} ($\epsilon$/$\eta$~Cha, $\sim$6 Myr; TW~Hya,
  $\sim$10 Myr; $\beta$~Pic, $\sim$12 Myr; Tucana-Horologium, $\sim$30
  Myr; Argus, $\sim$40 Myr; AB~Dor, $\sim$70 Myr). For consistency all
  temperatures were calculated from $V-I$ color and the transformation
  of \citet{Kenyon1995}. Lower main sequence members of IC~2391 with
  $V-I$ colors from \citet{Barrado-y-Navascues2004} are also plotted
  (filled squares, open squares denote upper limits). The decreasing
  trend in $EW$ with decreasing temperature in the older groups is
  readily apparent, as are the effects of lithium depletion with
  age. AP~Col is moderately lithium-depleted compared to the younger
  groups and lies between the $\beta$~Pic Lithium Depletion Boundary
  (LDB, dashed line defined by the two systems at $\sim$3300~K,
  $EW_{\rm Li I}\approx450$~m\AA) and the IC~2391 LDB (approximated by
  the dotted line). This constrains the age of AP~Col to between that
  of $\beta$~Pic (12 Myr) and IC~2391 \citep[$50 \pm
  5$~Myr][]{Barrado-y-Navascues2004}.\label{fig:lithium_groups} }
\end{figure}

Along with their kinematics and spatial positions, the lithium
distributions and color-magnitude diagrams of Argus and IC~2391 show
good agreement for the solar-type members of the \citet{Torres2008}
sample.  As already discussed, Figure~\ref{fig:lithium_groups} shows
the \citet{Barrado-y-Navascues2004} IC~2391 members that define the
LDB (approximated by the dotted line) at around M5 and an age of
50$\pm$5 Myr, subject to the inaccuracies of lithium dating
\citep{Yee2010,Song2002}. The position of AP~Col in this diagram is
consistent with an age similar to that of IC~2391. In fact, the star
appears to lie on the LDB, a position supported by the exact agreement
of its ($R-I$) color and that derived for the LDB by
\citet{Barrado-y-Navascues2004}.

In a wider context, \citet{da-Silva2009} found that Argus members have
a level of lithium depletion between that of the $\sim$30 Myr old
Tuc-Hor and the $\sim$70 Myr old AB~Dor associations. This is
consistent with the lithium age for IC~2391 above, and our putative
age range for AP~Col.  Thus, given its appropriate kinematics and age,
we claim that AP~Col, at 8.39 pc, is a likely member of the $\sim$40
Myr old Argus association.

\begin{figure}
\center
\includegraphics[angle=0,scale=.5]{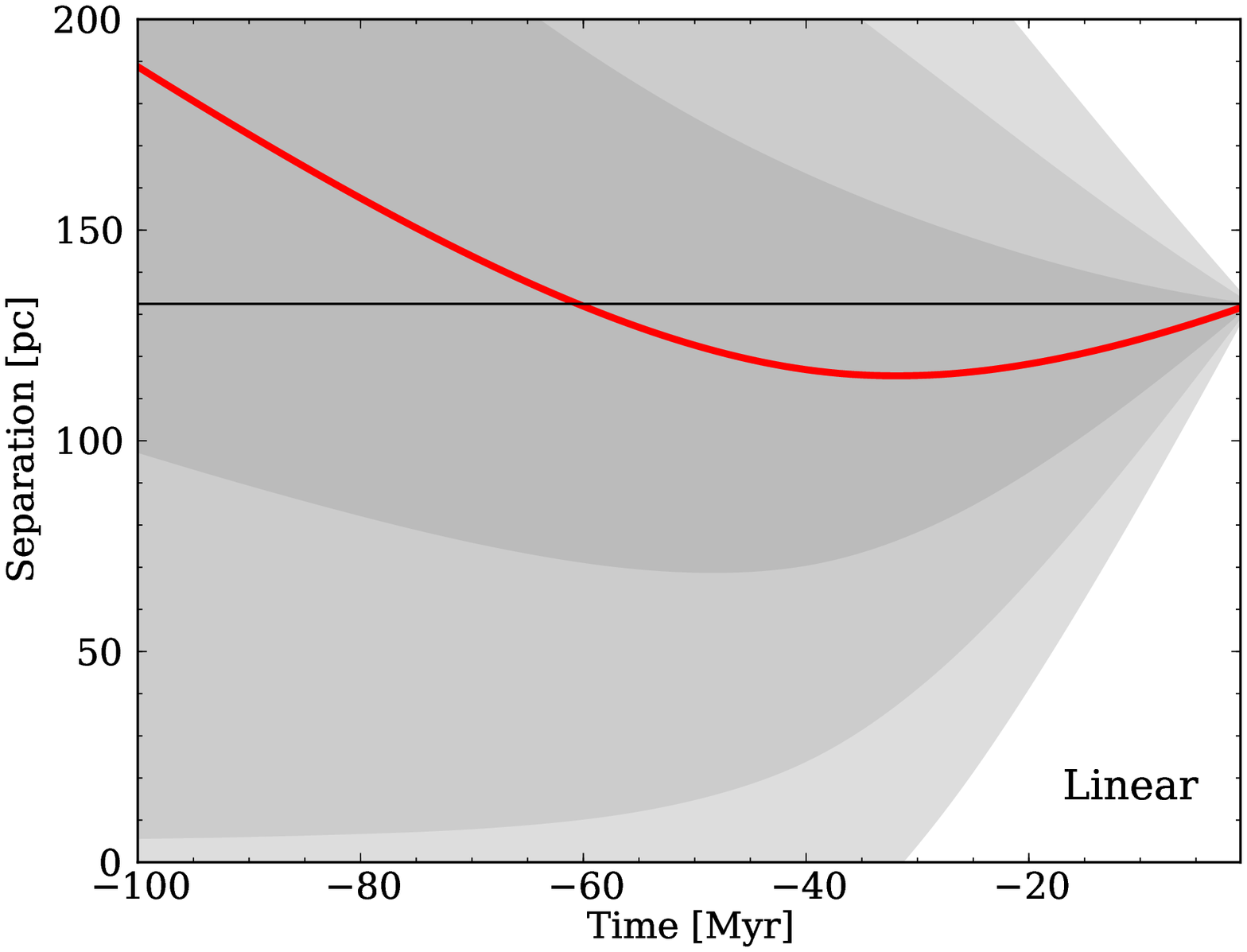}
\includegraphics[angle=0,scale=.5]{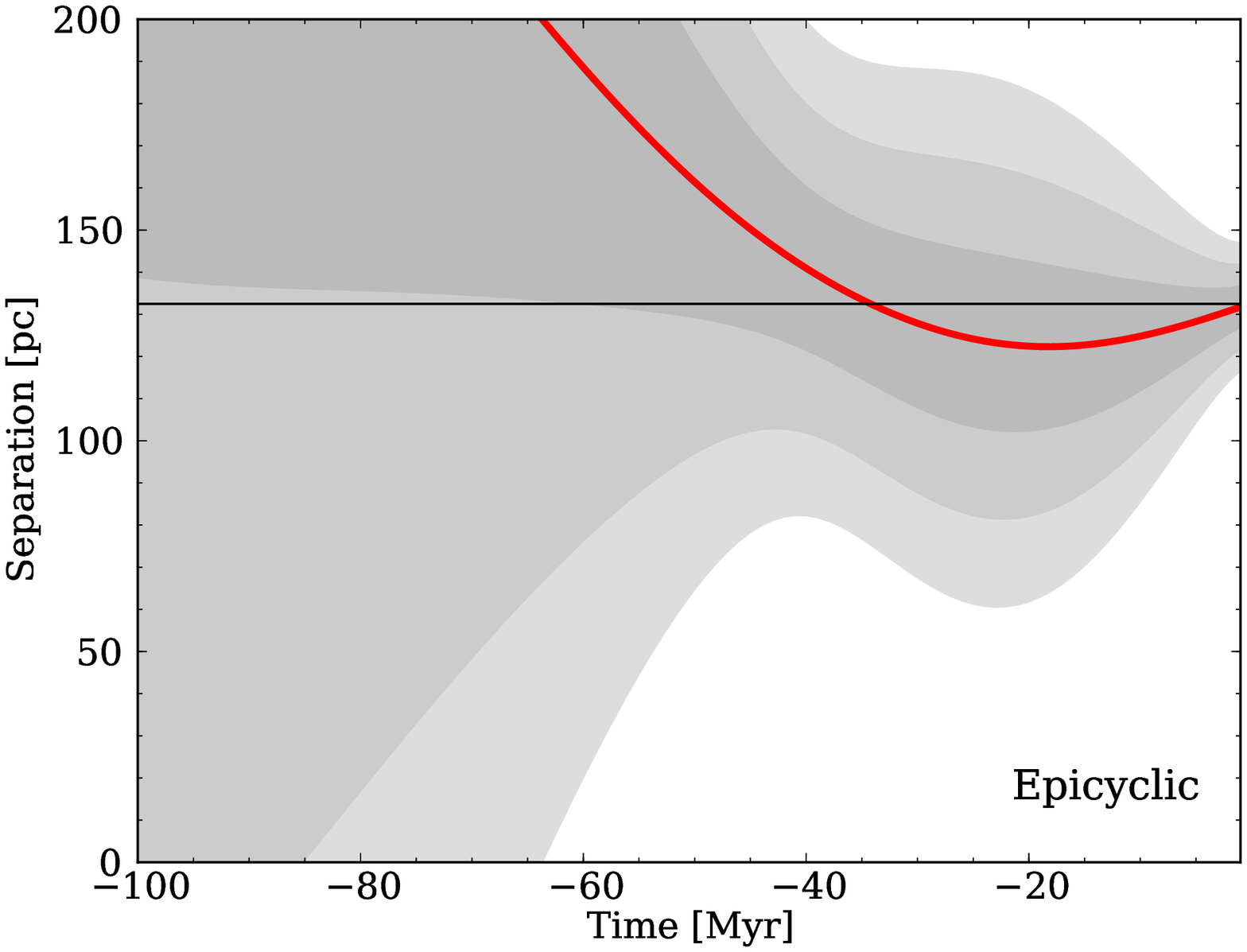}
\caption{The separation between AP~Col and the mean IC~2391 location
  (dotted line) as a function of time, with their current separation
  ($\sim$130 pc) on the right hand side.  Each plot shows the result
  of 1000 Monte Carlo simulations of the linear/ballistic (top) and
  epicyclic (bottom) approximations to galactic dynamics, using the
  observed uncertainties in the observed AP~Col and IC~2391 space
  motions. The shaded regions show 1 (darkest), 2 and 3$\sigma$
  (lightest) confidence intervals around the mean trend.  If AP~Col
  were a member of IC~2391, the separation of AP~Col and IC~2391
  should go to zero when they were both born ($\sim$40--50 Myr ago);
  they do not reasonably converge at the $\sim2\sigma$ level until
  $\sim$80 Myr ago.  This implies either interactions have modified
  the observed velocities, or that AP~Col did not form near the core
  of IC~2391.
  \label{fig:convergence}}
\end{figure}

One curious piece of evidence contradicts this scenario: if AP~Col is
an `evaporated' member of Argus/IC~2391, it should converge on the
mean location of IC~2391, roughly 40 Myr ago.  Using both linear
\citep{Murphy2010} and epicycle \citep{Makarov2004} approximations to
Galactic dynamics to retrace its motion, we find that AP~Col does not
converge with IC~2391 until (at the very earliest) 80 Myr ago (Figure
\ref{fig:convergence}).  This result suggests that either the epicycle
approximation is insufficient for a 40 Myr traceback, AP~Col (and
potentially Argus) have been kinematically perturbed since formation
(as the evaporated outer envelope of IC~2391), or AP~Col (and
potentially Argus) formed elsewhere at roughly the same time as
IC~2391.

While the evaporation and perturbation scenario can explain our
failure to trace AP~Col back to IC~2391, it does not explain why,
post-interaction(s), the velocity of AP~Col is still so close to that
of Argus/IC~2391.  We instead speculate, like \citet{Desidera2011},
that Argus is actually the product of gas \emph{surrounding} what
became IC~2391, thrust into star formation within a few million years
of the cluster, perhaps triggered by supernovae in IC~2391. A similar
relationship was suggested by \citet{Ortega2007} in connecting AB~Dor
with the Pleiades, though we have not connected them in this paper.

In any case, the physicality of Argus is beyond the purpose of this
paper; our main conclusion is that, as defined in \citet{Torres2008},
AP~Col is a member of the Argus association, and is now its closest
member.  With a presumed age of $\sim$40 Myr as part of the Argus
association, AP~Col is thus the second youngest member of the
immediate solar neighborhood, forming during the Eocene epoch on
Earth, and at 8.4 pc, the closest star with an age less than 100 Myr.

\vskip 0.5in
Acknowledgments: 

The RECONS effort is primarily supported by the National Science
Foundation through grants AST 05-07711 and AST 09-08402, and for many
years through NASA's Space Interferometry Mission.  Observations were
initially made possible by NOAO's Survey Program and have continued
via the SMARTS Consortium.

Some of the data presented herein were obtained at the W.M. Keck
Observatory, which is operated as a scientific partnership among the
California Institute of Technology, the University of California and
the National Aeronautics and Space Administration. The Observatory was
made possible by the generous financial support of the W.M. Keck
Foundation.  C.M. acknowledges support from the National Science
Foundation under AST-1003318.

This research has made use of results from the NStars project, NASA's
Astrophysics Data System Bibliographic Services, the SIMBAD and VizieR
databases operated at CDS Strasbourg, France, the SuperCOSMOS Sky
Survey, the 2MASS database, and the Keck Observatory Archive (KOA),
which is operated by the W. M. Keck Observatory, and the NASA Exoplanet
Science Institute (NExScI), under contract with the National
Aeronautics and Space Administration.

The authors also wish to thank the following people for their
assistance:

The rest of the RECONS group, and the staff of the Cerro Tololo
Inter-american Observatory (particularly E. Cosgrove, A. Gomez,
A. Miranda and J. Vasquez)

M.S. Bessell for the SSO WiFeS low-resolution spectroscopy.

W. Hartkopf and B. Mason for the USNO Speckle results.

S. Vogt and P. Butler for obtaining the HIRES spectrum of AP~Col for
us.

R.J. White and E.E. Mamajek for constructive comments on youth and
kinematics.

The anonymous referee for his or her suggestions.


\bibliographystyle{apj}
\bibliography{riedel_a}

\begin{thebibliography}{71}
\expandafter\ifx\csname natexlab\endcsname\relax\def\natexlab#1{#1}\fi

\bibitem[{{Ball} \& {Bromage}(1995)}]{Ball1995}
{Ball}, B., \& {Bromage}, G. 1995, in Lecture Notes in Physics, Berlin Springer
  Verlag, Vol. 454, IAU Colloq. 151: Flares and Flashes, ed. {J.~Greiner,
  H.~W.~Duerbeck, \& R.~E.~Gershberg}, 67--+

\bibitem[{{Barrado y Navascu{\'e}s} {et~al.}(2004){Barrado y Navascu{\'e}s},
  {Stauffer}, \& {Jayawardhana}}]{Barrado-y-Navascues2004}
{Barrado y Navascu{\'e}s}, D., {Stauffer}, J.~R., \& {Jayawardhana}, R. 2004,
  \apj, 614, 386

\bibitem[{{Barrado y Navascu{\'e}s} {et~al.}(1999){Barrado y Navascu{\'e}s},
  {Stauffer}, {Song}, \& {Caillault}}]{Barrado-y-Navascues1999}
{Barrado y Navascu{\'e}s}, D., {Stauffer}, J.~R., {Song}, I., \& {Caillault},
  J. 1999, \apjl, 520, L123

\bibitem[{{Bergfors} {et~al.}(2010){Bergfors}, {Brandner}, {Janson}, {Daemgen},
  {Geissler}, {Henning}, {Hippler}, {Hormuth}, {Joergens}, \&
  {K{\"o}hler}}]{Bergfors2010}
{Bergfors}, C., {et~al.} 2010, \aap, 520, A54+

\bibitem[{{Biller} \& {Close}(2007)}]{Biller2007}
{Biller}, B.~A., \& {Close}, L.~M. 2007, \apjl, 669, L41

\bibitem[{{Cutri} {et~al.}(2003){Cutri}, {Skrutskie}, {van Dyk}, {Beichman},
  {Carpenter}, {Chester}, {Cambresy}, {Evans}, {Fowler}, {Gizis}, {Howard},
  {Huchra}, {Jarrett}, {Kopan}, {Kirkpatrick}, {Light}, {Marsh}, {McCallon},
  {Schneider}, {Stiening}, {Sykes}, {Weinberg}, {Wheaton}, {Wheelock}, \&
  {Zacarias}}]{Cutri2003}
{Cutri}, R.~M., {et~al.} 2003, {2MASS All Sky Catalog of point sources.}, ed.
  {Cutri, R.~M., Skrutskie, M.~F., van Dyk, S., Beichman, C.~A., Carpenter,
  J.~M., Chester, T., Cambresy, L., Evans, T., Fowler, J., Gizis, J., Howard,
  E., Huchra, J., Jarrett, T., Kopan, E.~L., Kirkpatrick, J.~D., Light, R.~M.,
  Marsh, K.~A., McCallon, H., Schneider, S., Stiening, R., Sykes, M., Weinberg,
  M., Wheaton, W.~A., Wheelock, S., \& Zacarias, N.}

\bibitem[{{da Silva} {et~al.}(2009){da Silva}, {Torres}, {de La Reza}, {Quast},
  {Melo}, \& {Sterzik}}]{da-Silva2009}
{da Silva}, L., {Torres}, C.~A.~O., {de La Reza}, R., {Quast}, G.~R., {Melo},
  C.~H.~F., \& {Sterzik}, M.~F. 2009, \aap, 508, 833

\bibitem[{{de la Reza} {et~al.}(1989){de la Reza}, {Torres}, {Quast},
  {Castilho}, \& {Vieira}}]{de-la-Reza1989}
{de la Reza}, R., {Torres}, C.~A.~O., {Quast}, G., {Castilho}, B.~V., \&
  {Vieira}, G.~L. 1989, \apjl, 343, L61

\bibitem[{{Delfosse} {et~al.}(1999){Delfosse}, {Forveille}, {Beuzit}, {Udry},
  {Mayor}, \& {Perrier}}]{Delfosse1999}
{Delfosse}, X., {Forveille}, T., {Beuzit}, J., {Udry}, S., {Mayor}, M., \&
  {Perrier}, C. 1999, \aap, 344, 897

\bibitem[{{Desidera} {et~al.}(2011){Desidera}, {Covino}, {Messina}, {D'Orazi
  D'Orazi}, {Alcal{\'a}}, {Brugaletta}, {Carson}, {Lanzafame}, \&
  {Launhardt}}]{Desidera2011}
{Desidera}, S., {et~al.} 2011, \aap, 529, A54+

\bibitem[{{Dopita} {et~al.}(2007){Dopita}, {Hart}, {McGregor}, {Oates},
  {Bloxham}, \& {Jones}}]{Dopita2007}
{Dopita}, M., {Hart}, J., {McGregor}, P., {Oates}, P., {Bloxham}, G., \&
  {Jones}, D. 2007, \apss, 310, 255

\bibitem[{{Dopita} {et~al.}(2010){Dopita}, {Rhee}, {Farage}, {McGregor},
  {Bloxham}, {Green}, {Roberts}, {Neilson}, {Wilson}, {Young}, {Firth},
  {Busarello}, \& {Merluzzi}}]{Dopita2010}
{Dopita}, M., {et~al.} 2010, \apss, 327, 245

\bibitem[{{Ducourant} {et~al.}(2008){Ducourant}, {Teixeira}, {Chauvin},
  {Daigne}, {Le Campion}, {Song}, \& {Zuckerman}}]{Ducourant2008}
{Ducourant}, C., {Teixeira}, R., {Chauvin}, G., {Daigne}, G., {Le Campion},
  J.-F., {Song}, I., \& {Zuckerman}, B. 2008, \aap, 477, L1

\bibitem[{{Eggen}(1958)}]{Eggen1958}
{Eggen}, O.~J. 1958, \mnras, 118, 65

\bibitem[{{Famaey} {et~al.}(2008){Famaey}, {Siebert}, \&
  {Jorissen}}]{Famey2008}
{Famaey}, B., {Siebert}, A., \& {Jorissen}, A. 2008, \aap, 483, 453

\bibitem[{{Gizis} {et~al.}(2007){Gizis}, {Jao}, {Subasavage}, \&
  {Henry}}]{Gizis2007}
{Gizis}, J.~E., {Jao}, W., {Subasavage}, J.~P., \& {Henry}, T.~J. 2007, \apjl,
  669, L45

\bibitem[{{Graham}(1982)}]{Graham1982}
{Graham}, J.~A. 1982, \pasp, 94, 244

\bibitem[{{Haisch} {et~al.}(2005){Haisch}, {Jayawardhana}, \&
  {Alves}}]{Haisch2005}
{Haisch}, Jr., K.~E., {Jayawardhana}, R., \& {Alves}, J. 2005, \apjl, 627, L57

\bibitem[{{Haisch} {et~al.}(2001){Haisch}, {Lada}, \& {Lada}}]{Haisch2001}
{Haisch}, Jr., K.~E., {Lada}, E.~A., \& {Lada}, C.~J. 2001, \apjl, 553, L153

\bibitem[{{Hambly} {et~al.}(2001){Hambly}, {MacGillivray}, {Read}, {Tritton},
  {Thomson}, {Kelly}, {Morgan}, {Smith}, {Driver}, {Williamson}, {Parker},
  {Hawkins}, {Williams}, \& {Lawrence}}]{Hambly2001}
{Hambly}, N.~C., {et~al.} 2001, \mnras, 326, 1279

\bibitem[{{Hawley} {et~al.}(1996){Hawley}, {Gizis}, \& {Reid}}]{Hawley1996}
{Hawley}, S.~L., {Gizis}, J.~E., \& {Reid}, I.~N. 1996, \aj, 112, 2799

\bibitem[{{Hayashi}(1966)}]{Hayashi1966}
{Hayashi}, C. 1966, \araa, 4, 171

\bibitem[{{Henry} {et~al.}(1999){Henry}, {Franz}, {Wasserman}, {Benedict},
  {Shelus}, {Ianna}, {Kirkpatrick}, \& {McCarthy}}]{Henry1999}
{Henry}, T.~J., {Franz}, O.~G., {Wasserman}, L.~H., {Benedict}, G.~F.,
  {Shelus}, P.~J., {Ianna}, P.~A., {Kirkpatrick}, J.~D., \& {McCarthy}, Jr.,
  D.~W. 1999, \apj, 512, 864

\bibitem[{{Henry} {et~al.}(2006){Henry}, {Jao}, {Subasavage}, {Beaulieu},
  {Ianna}, {Costa}, \& {M{\'e}ndez}}]{Henry2006}
{Henry}, T.~J., {Jao}, W., {Subasavage}, J.~P., {Beaulieu}, T.~D., {Ianna},
  P.~A., {Costa}, E., \& {M{\'e}ndez}, R.~A. 2006, \aj, 132, 2360

\bibitem[{{Henry} {et~al.}(2004){Henry}, {Subasavage}, {Brown}, {Beaulieu},
  {Jao}, \& {Hambly}}]{Henry2004}
{Henry}, T.~J., {Subasavage}, J.~P., {Brown}, M.~A., {Beaulieu}, T.~D., {Jao},
  W., \& {Hambly}, N.~C. 2004, \aj, 128, 2460

\bibitem[{{Henry} {et~al.}(2002){Henry}, {Walkowicz}, {Barto}, \&
  {Golimowski}}]{Henry2002}
{Henry}, T.~J., {Walkowicz}, L.~M., {Barto}, T.~C., \& {Golimowski}, D.~A.
  2002, \aj, 123, 2002

\bibitem[{{Jao} {et~al.}(2005){Jao}, {Henry}, {Subasavage}, {Brown}, {Ianna},
  {Bartlett}, {Costa}, \& {M{\'e}ndez}}]{Jao2005}
{Jao}, W., {Henry}, T.~J., {Subasavage}, J.~P., {Brown}, M.~A., {Ianna}, P.~A.,
  {Bartlett}, J.~L., {Costa}, E., \& {M{\'e}ndez}, R.~A. 2005, \aj, 129, 1954

\bibitem[{{Jao} {et~al.}(2011){Jao}, {Henry}, {Subasavage}, {Winters},
  {Riedel}, \& {Ianna}}]{Jao2011}
{Jao}, W., {Henry}, T.~J., {Subasavage}, J.~P., {Winters}, J.~G., {Riedel},
  A.~R., \& {Ianna}, P.~A. 2011, \aj, 141, 117

\bibitem[{{Jeffries} \& {Naylor}(2001)}]{Jeffries2001}
{Jeffries}, R.~D., \& {Naylor}, T. 2001, in Astronomical Society of the Pacific
  Conference Series, Vol. 243, From Darkness to Light: Origin and Evolution of
  Young Stellar Clusters, ed. {T.~Montmerle \& P.~Andr{\'e}}, 633--+

\bibitem[{{Kenyon} \& {Hartmann}(1995)}]{Kenyon1995}
{Kenyon}, S.~J., \& {Hartmann}, L. 1995, \apjs, 101, 117

\bibitem[{{Kharchenko}(2001)}]{Kharchenko2001}
{Kharchenko}, N.~V. 2001, Kinematika i Fizika Nebesnykh Tel, 17, 409

\bibitem[{{Kirkpatrick} {et~al.}(1991){Kirkpatrick}, {Henry}, \&
  {McCarthy}}]{Kirkpatrick1991}
{Kirkpatrick}, J.~D., {Henry}, T.~J., \& {McCarthy}, Jr., D.~W. 1991, \apjs,
  77, 417

\bibitem[{{Landolt}(1992)}]{Landolt1992}
{Landolt}, A.~U. 1992, \aj, 104, 340

\bibitem[{{Landolt}(2007)}]{Landolt2007}
---. 2007, \aj, 133, 2502

\bibitem[{{Lawson} {et~al.}(2009){Lawson}, {Lyo}, \& {Bessell}}]{Lawson2009b}
{Lawson}, W.~A., {Lyo}, A., \& {Bessell}, M.~S. 2009, \mnras, 400, L29

\bibitem[{{L{\'o}pez-Santiago} {et~al.}(2009){L{\'o}pez-Santiago}, {Micela}, \&
  {Montes}}]{Lopez-Santiago2009}
{L{\'o}pez-Santiago}, J., {Micela}, G., \& {Montes}, D. 2009, \aap, 499, 129

\bibitem[{{Lyo} {et~al.}(2004){Lyo}, {Lawson}, \& {Bessell}}]{Lyo2004a}
{Lyo}, A.-R., {Lawson}, W.~A., \& {Bessell}, M.~S. 2004, \mnras, 355, 363

\bibitem[{{Makarov} {et~al.}(2004){Makarov}, {Olling}, \&
  {Teuben}}]{Makarov2004}
{Makarov}, V.~V., {Olling}, R.~P., \& {Teuben}, P.~J. 2004, \mnras, 352, 1199

\bibitem[{{Makarov} \& {Urban}(2000)}]{Makarov2000}
{Makarov}, V.~V., \& {Urban}, S. 2000, \mnras, 317, 289

\bibitem[{{Mamajek}(2009)}]{Mamajek2009}
{Mamajek}, E.~E. 2009, in IAU Symposium, Vol. 258, IAU Symposium, ed.
  {E.~E.~Mamajek, D.~R.~Soderblom, \& R.~F.~G.~Wyse}, 375--382

\bibitem[{{Murphy} {et~al.}(2010){Murphy}, {Lawson}, \& {Bessell}}]{Murphy2010}
{Murphy}, S.~J., {Lawson}, W.~A., \& {Bessell}, M.~S. 2010, \mnras, 406, L50

\bibitem[{{Nidever} {et~al.}(2002){Nidever}, {Marcy}, {Butler}, {Fischer}, \&
  {Vogt}}]{Nidever2002}
{Nidever}, D.~L., {Marcy}, G.~W., {Butler}, R.~P., {Fischer}, D.~A., \& {Vogt},
  S.~S. 2002, \apjs, 141, 503

\bibitem[{{Ortega} {et~al.}(2007){Ortega}, {Jilinski}, {de La Reza}, \&
  {Bazzanella}}]{Ortega2007}
{Ortega}, V.~G., {Jilinski}, E., {de La Reza}, R., \& {Bazzanella}, B. 2007,
  \mnras, 377, 441

\bibitem[{{Platais} {et~al.}(2007){Platais}, {Melo}, {Mermilliod},
  {Kozhurina-Platais}, {Fulbright}, {M{\'e}ndez}, {Altmann}, \&
  {Sperauskas}}]{Platais2007}
{Platais}, I., {Melo}, C., {Mermilliod}, J., {Kozhurina-Platais}, V.,
  {Fulbright}, J.~P., {M{\'e}ndez}, R.~A., {Altmann}, M., \& {Sperauskas}, J.
  2007, \aap, 461, 509

\bibitem[{{Pojmanski}(1997)}]{Pojmanski1997}
{Pojmanski}, G. 1997, \actaa, 47, 467

\bibitem[{{Reid} \& {Hawley}(1999)}]{Reid1999}
{Reid}, I.~N., \& {Hawley}, S.~L. 1999, \aj, 117, 343

\bibitem[{{Reiners} {et~al.}(2009){Reiners}, {Basri}, \&
  {Browning}}]{Reiners2009}
{Reiners}, A., {Basri}, G., \& {Browning}, M. 2009, \apj, 692, 538

\bibitem[{{Riaz} {et~al.}(2006){Riaz}, {Gizis}, \& {Harvin}}]{Riaz2006}
{Riaz}, B., {Gizis}, J.~E., \& {Harvin}, J. 2006, \aj, 132, 866

\bibitem[{{Riedel} {et~al.}(2010){Riedel}, {Subasavage}, {Finch}, {Jao},
  {Henry}, {Winters}, {Brown}, {Ianna}, {Costa}, \& {Mendez}}]{Riedel2010}
{Riedel}, A.~R., {et~al.} 2010, \aj, 140, 897

\bibitem[{{Robrade} {et~al.}(2004){Robrade}, {Ness}, \&
  {Schmitt}}]{Robrade2004}
{Robrade}, J., {Ness}, J., \& {Schmitt}, J.~H.~M.~M. 2004, \aap, 413, 317

\bibitem[{{Rodriguez} {et~al.}(2011){Rodriguez}, {Bessell}, {Zuckerman}, \&
  {Kastner}}]{Rodriguez2011}
{Rodriguez}, D.~R., {Bessell}, M.~S., {Zuckerman}, B., \& {Kastner}, J.~H.
  2011, \apj, 727, 62

\bibitem[{{Scholz} {et~al.}(2005){Scholz}, {Lo Curto}, {M{\'e}ndez},
  {Hambaryan}, {Costa}, {Henry}, \& {Schwope}}]{Scholz2005}
{Scholz}, R., {Lo Curto}, G., {M{\'e}ndez}, R.~A., {Hambaryan}, V., {Costa},
  E., {Henry}, T.~J., \& {Schwope}, A.~D. 2005, \aap, 439, 1127

\bibitem[{{Slesnick} {et~al.}(2006{\natexlab{a}}){Slesnick}, {Carpenter}, \&
  {Hillenbrand}}]{Slesnick2006}
{Slesnick}, C.~L., {Carpenter}, J.~M., \& {Hillenbrand}, L.~A.
  2006{\natexlab{a}}, \aj, 131, 3016

\bibitem[{{Slesnick} {et~al.}(2006{\natexlab{b}}){Slesnick}, {Carpenter},
  {Hillenbrand}, \& {Mamajek}}]{Slesnick2006b}
{Slesnick}, C.~L., {Carpenter}, J.~M., {Hillenbrand}, L.~A., \& {Mamajek},
  E.~E. 2006{\natexlab{b}}, \aj, 132, 2665

\bibitem[{{Soderblom} {et~al.}(2005){Soderblom}, {Nelan}, {Benedict},
  {McArthur}, {Ramirez}, {Spiesman}, \& {Jones}}]{Soderblom2005}
{Soderblom}, D.~R., {Nelan}, E., {Benedict}, G.~F., {McArthur}, B., {Ramirez},
  I., {Spiesman}, W., \& {Jones}, B.~F. 2005, \aj, 129, 1616

\bibitem[{{Song} {et~al.}(2002){Song}, {Bessell}, \& {Zuckerman}}]{Song2002}
{Song}, I., {Bessell}, M.~S., \& {Zuckerman}, B. 2002, \apjl, 581, L43

\bibitem[{{Stauffer} {et~al.}(1998){Stauffer}, {Schultz}, \&
  {Kirkpatrick}}]{Stauffer1998}
{Stauffer}, J.~R., {Schultz}, G., \& {Kirkpatrick}, J.~D. 1998, \apjl, 499,
  L199+

\bibitem[{{Subasavage} {et~al.}(2009){Subasavage}, {Jao}, {Henry}, {Bergeron},
  {Dufour}, {Ianna}, {Costa}, \& {M{\'e}ndez}}]{Subasavage2009}
{Subasavage}, J.~P., {Jao}, W., {Henry}, T.~J., {Bergeron}, P., {Dufour}, P.,
  {Ianna}, P.~A., {Costa}, E., \& {M{\'e}ndez}, R.~A. 2009, \aj, 137, 4547

\bibitem[{{Teixeira} {et~al.}(2009){Teixeira}, {Ducourant}, {Chauvin},
  {Krone-Martins}, {Bonnefoy}, \& {Song}}]{Teixeira2009}
{Teixeira}, R., {Ducourant}, C., {Chauvin}, G., {Krone-Martins}, A.,
  {Bonnefoy}, M., \& {Song}, I. 2009, \aap, 503, 281

\bibitem[{{Torres} {et~al.}(2003){Torres}, {Quast}, {de La Reza}, {da Silva},
  {Melo}, \& {Sterzik}}]{Torres2003}
{Torres}, C.~A.~O., {Quast}, G.~R., {de La Reza}, R., {da Silva}, L., {Melo},
  C.~H.~F., \& {Sterzik}, M. 2003, in Astrophysics and Space Science Library,
  Vol. 299, Open Issues in Local Star Formation, ed. {J.~L{\'e}pine \&
  J.~Gregorio-Hetem}, 83--+

\bibitem[{{Torres} {et~al.}(2008){Torres}, {Quast}, {Melo}, \&
  {Sterzik}}]{Torres2008}
{Torres}, C.~A.~O., {Quast}, G.~R., {Melo}, C.~H.~F., \& {Sterzik}, M.~F. 2008,
  Handbook of Star Forming Regions, Volume II (ASP Press), 757

\bibitem[{{Trumpler}(1921)}]{Trumpler1921}
{Trumpler}, R.~J. 1921, Lick Observatory Bulletin, 10, 110

\bibitem[{{van Altena} {et~al.}(1995){van Altena}, {Lee}, \&
  {Hoffleit}}]{van-Altena1995}
{van Altena}, W.~F., {Lee}, J.~T., \& {Hoffleit}, E.~D. 1995, {The general
  catalogue of trigonometric [stellar] parallaxes}, ed. {van Altena, W.~F.,
  Lee, J.~T., \& Hoffleit, E.~D.}

\bibitem[{{van Leeuwen}(2007)}]{van-Leeuwen2007}
{van Leeuwen}, F., ed. 2007, Astrophysics and Space Science Library, Vol. 350,
  {Hipparcos, the New Reduction of the Raw Data}

\bibitem[{{Vogt}(1987)}]{Vogt1987}
{Vogt}, S.~S. 1987, \pasp, 99, 1214

\bibitem[{{Vogt} {et~al.}(1994){Vogt}, {Allen}, {Bigelow}, {Bresee}, {Brown},
  {Cantrall}, {Conrad}, {Couture}, {Delaney}, {Epps}, {Hilyard}, {Hilyard},
  {Horn}, {Jern}, {Kanto}, {Keane}, {Kibrick}, {Lewis}, {Osborne},
  {Pardeilhan}, {Pfister}, {Ricketts}, {Robinson}, {Stover}, {Tucker}, {Ward},
  \& {Wei}}]{Vogt1994}
{Vogt}, S.~S., {et~al.} 1994, in Society of Photo-Optical Instrumentation
  Engineers (SPIE) Conference Series, Vol. 2198, Society of Photo-Optical
  Instrumentation Engineers (SPIE) Conference Series, ed. {D.~L.~Crawford \&
  E.~R.~Craine}, 362--+

\bibitem[{{Weinberger} {et~al.}(2011){Weinberger}, {Anglada-Escud{\'e}}, \&
  {Boss}}]{Weinberger2011}
{Weinberger}, A.~J., {Anglada-Escud{\'e}}, G., \& {Boss}, A. 2011, in Bulletin
  of the American Astronomical Society, Vol.~43, American Astronomical Society
  Meeting Abstracts \#217, \#340.12

\bibitem[{{Winters} {et~al.}(2011){Winters}, {Henry}, {Jao}, {Subasavage},
  {Finch}, \& {Hambly}}]{Winters2011}
{Winters}, J.~G., {Henry}, T.~J., {Jao}, W., {Subasavage}, J.~P., {Finch},
  C.~T., \& {Hambly}, N.~C. 2011, \aj, 141, 21

\bibitem[{{Yee} \& {Jensen}(2010)}]{Yee2010}
{Yee}, J.~C., \& {Jensen}, E.~L.~N. 2010, \apj, 711, 303

\bibitem[{{Zuckerman} {et~al.}(2011){Zuckerman}, {Rhee}, {Song}, \&
  {Bessell}}]{Zuckerman2011}
{Zuckerman}, B., {Rhee}, J.~H., {Song}, I., \& {Bessell}, M.~S. 2011, \apj,
  732, 61

\bibitem[{{Zuckerman} \& {Song}(2004)}]{Zuckerman2004}
{Zuckerman}, B., \& {Song}, I. 2004, \araa, 42, 685

\end{thebibliography}

\clearpage
\voffset60pt{
\begin{deluxetable}{lcccccccccc}
\rotate
\tabletypesize{\scriptsize}
\tablecolumns{11} 
\tablewidth{0pt}
\tablecaption{AP~Col Spectroscopic Observations Summary \label{tab:optspec}}
\tablehead{
 \colhead{UT Date} & 
 \colhead{Instrument} & 
 \colhead{Setup} & 
 \colhead{Coverage} & 
 \colhead{Resolving\tablenotemark{a}} & 
 \colhead{S/N} & 
 \colhead{$\lambda$ of S/N\tablenotemark{b}} &
 \colhead{H$\alpha$ EW} &
 \colhead{Li $\lambda$6708 EW} &
 \colhead{RV\tablenotemark{c}}  &
 \colhead{{\it vsini}} \\
 
 \colhead{} &
 \colhead{} &
 \colhead{} &
 \colhead{\AA} &
 \colhead{Power} &
 \colhead{} &
 \colhead{\AA} &
 \colhead{\AA} &
 \colhead{\AA} &
 \colhead{km s$^{-1}$} &
 \colhead{km s$^{-1}$}
}
\startdata
 08 Jan 2011 & WiFeS    & R$_{7000}$/RT480\tablenotemark{d}& 5500-7000 & 7,000 & 30 & 6300 & -7.5$\pm$1.0 & 0.25$\pm$0.1  & 21.3$\pm$1.0 & \\
 08 Jan 2011 & WiFeS    & R$_{7000}$/RT480   & 5500-7000     &  7,000 &  25 & 6300 &  -8.0$\pm$1.0 & 0.25$\pm$0.1  & 24.1$\pm$0.8 & \\
 25 Jan 2011 & Hamilton & 800 $\mu$m slit, Dewar \#6 & 3850-8850 & 40,000 & 25 & 6700 & -35$\pm$3  & 0.19$\pm$0.03 & 23 $\pm$1  & \\
 25 Jan 2011 & WiFeS    & B/R$_{3000}$/RT560 & 3400-9650     &  3,000 &  60 & 7400 & -28 $\pm$ 3   &              &            & \\
 25 Jan 2011 & WiFeS    & B/R$_{3000}$/RT560 & 3400-9650     &  3,000 &  60 & 7400 & -26 $\pm$ 3   &              &            & \\
 25 Jan 2011 & WiFeS    & B/R$_{3000}$/RT560 & 3400-9650     &  3,000 &  60 & 7400 & -35 $\pm$ 3   &              &            & \\
 26 Jan 2011 & WiFeS    & B/R$_{3000}$/RT560 & 3400-9650     &  3,000 &  60 & 7400 & -12 $\pm$ 3   &              &            & \\
 11 Feb 2011 & WiFeS    & R$_{7000}$/RT480   & 5500-7000     &  7,000 &  30 & 6300 & -13.5$\pm$1.0 & 0.25$\pm$0.05 & 21.5$\pm$0.9 & \\
 24 Feb 2011 & WiFeS    & R$_{7000}$/RT480   & 5500-7000     &  7,000 &  25 & 6300 &  -7.5$\pm$1.0 & 0.3 $\pm$0.05 & 20.0$\pm$0.8 & \\
 16 Mar 2011 & WiFeS    & R$_{7000}$/RT480   & 5500-7000     &  7,000 &  35 & 6300 & -12.1$\pm$1.0 & 0.25$\pm$0.1  & 24.1$\pm$1.0 & \\
 17 Mar 2011 & WiFeS    & R$_{7000}$/RT480   & 5500-7000     &  7,000 &  40 & 6300 &  -6.5$\pm$1.0 & 0.25$\pm$0.05 & 23.2$\pm$1.0 & \\
 17 Mar 2011 & WiFeS    & R$_{7000}$/RT480   & 5500-7000     &  7,000 &  50 & 6300 &  -6.0$\pm$1.0 & 0.3 $\pm$0.05 & 22.5$\pm$1.1 & \\
 17 Mar 2011 & HIRES    & Red Collimator     & 3580-7950     & 50,000 &  20 & 6700 &  -7.3$\pm$0.5 & 0.37$\pm$0.03 & 22.3$\pm$0.3 & 11$\pm$1 \\
\enddata
\tablenotetext{a}{Resolution is measured from the FWHM of single
arclines in our comparison spectra.}
\tablenotetext{b}{Wavelength where S/N measurement is made in the
spectrum.}
\tablenotetext{c}{RV errors for WiFeS are internal errors with
reference to the standards.  The approximate error per epoch is 2 km
s$^{-1}$}
\tablenotetext{d}{RT480 and RT560 are dichroics for the beam splitter}
\end{deluxetable}
}


\clearpage

\thispagestyle{empty}

\voffset010pt{
\begin{deluxetable}{lll}
\setlength{\tabcolsep}{0.04in}
\tablewidth{0pt}
\tabletypesize{\scriptsize}
\tablecaption{AP~Col Vital Parameters\label{tab:properties}}
\tablehead{\colhead{}           &
           \colhead{AP~Col}
}
\startdata
\hline
Position (J2000 E2000)  & 06:04:52.16 $-$34:33:36.0\\
\hline
\multicolumn{2}{c}{Astrometric} \\
\hline
$\pi_{rel}$ (mas)       & 118.26$\pm$0.97\\
$\pi_{corr}$ (mas)      & 0.95$\pm$0.11\\
$\pi_{abs}$ (mas)       & 119.21$\pm$0.98\\
Distance (pc)           & 8.39$\pm$0.07 \\
$\mu_{\alpha,\delta}$ (mas/yr) & (27.33, 340.92) $\pm$ (0.35)\\
$\mu$ (mas/yr)          & 342.0$\pm$0.5 \\
P.A. (deg)              & 004.6$\pm$0.13 \\
V$_{tan}$ (km s$^{-1}$) & 13.60$\pm$0.11 \\
\hline
\multicolumn{2}{c}{Photometric} \\
\hline
$V_J$                   & 12.96$\pm$0.01 \\
$R_{KC}$                & 11.49$\pm$0.02 \\
$I_{KC}$                &  9.60$\pm$0.01 \\
$J_{2MASS}$             &  7.74$\pm$0.03 \\
$H_{2MASS}$             &  7.18$\pm$0.02 \\
$K_{s 2MASS}$           &  6.87$\pm$0.02 \\
$M_V$                   & 13.34 \\
$V_J - K_{s 2MASS}$     &  6.09 \\
$L_{x} / L_{bol}$       &  -2.95$\pm$0.16 \\
log($L_{x}$)            &  28.49$\pm$17\% \\
$Variability$ (mag)     & 0.017 ($V_J$)\\
\hline
\multicolumn{2}{c}{Spectroscopic} \\
\hline
Spectral Type           & M4.5e\tablenotemark{a} \\
Li {\sc I} $\lambda$6708 EW (\AA)& $0.28\pm0.02$ \\
$H\alpha$ EW (\AA)      & $-$9.1$\pm$5.2 [variable $-$6 to $-$35]\\
V$_{rad}$ (km s$^{-1}$) & +22.4$\pm$0.3 \\
v$sin$i (km s$^{-1}$)   & 11$\pm$1 \\
\hline
\multicolumn{2}{c}{Derived Quantities} \\
\hline
$X$ (pc)                & $-$3.72$\pm$0.04 \\
$Y$ (pc)                & $-$6.70$\pm$0.08 \\
$Z$ (pc)                & $-$3.41$\pm$0.04 \\
$U$ (km s$^{-1}$)       & $-$21.98$\pm$0.17 \\
$V$ (km s$^{-1}$)       & $-$13.58$\pm$0.24 \\
$W$ (km s$^{-1}$)       & $-$4.45$\pm$0.13 \\
Isochronal Age (Myr)    & 12--70 \\
Na {\sc I} (gravity) Age (Myr) & 12--100 \\
Li {\sc I} Age (Myr)    & 12--50 \\
\hline
\enddata
\tablenotetext{a}{Measured type is M4.5Ve, but AP~Col is not a main
sequence star.}
\vfil\eject

\end{deluxetable}
}

\clearpage

\begin{deluxetable}{lcc}
\tabletypesize{\normalsize}
\tablecolumns{3}
\tablewidth{0pt}
\tablecaption{Emission Lines Detected in HIRES AP~Col Optical Spectra \label{tab:emlines}}
\tablehead{
 \colhead{Transition} &
 \colhead{Rest Wavelength} &
 \colhead{EW} \\
 \colhead{} &
 \colhead{(\AA\ in air )} &
 \colhead{(\AA )}
}
\startdata
H$\alpha$   & 6562.852 &  $-$7.3$\pm$0.5 \\
Na D$_1$    & 5895.924 &  $-$0.4$\pm$0.1\tablenotemark{a} \\
Na D$_2$    & 5889.951 &  $-$1.1$\pm$0.1\tablenotemark{a} \\
He\,I       & 5875.621 &  $-$0.8$\pm$0.1\tablenotemark{a} \\
H$\beta$    & 4861.350 &  $-$8$\pm$1 \\
He\,I       & 4471.480 &  $-$0.3$\pm$0.1 \\
H$\gamma$   & 4340.472 &  $-$6$\pm$1 \\
H$\delta$   & 4101.734 &  $-$5$\pm$1 \\
H$\epsilon$ & 3970.075 &  $-$4$\pm$1 \\
Ca\,II H    & 3968.470 & $-$10$\pm$2 \\
Ca\,II K    & 3933.660 & $-$13$\pm$3 \\
H8          & 3889.064 &  $-$5$\pm$2 \\
H9          & 3835.397 &  $-$4$\pm$2\tablenotemark{b} \\
H10         & 3797.909 &  $-$0.7$\pm$0.2\tablenotemark{b} \\
\enddata

\tablenotetext{a}{~Line contaminated by iodine absorption}
\tablenotetext{b}{Continuum around line not significantly detected}
\end{deluxetable}


\clearpage

\begin{deluxetable}{lccccccccc}
\rotate
\tabletypesize{\normalsize}
\tablecolumns{10}
\tablewidth{0pt}
\tablecaption{Polynomial coefficients for empirical association isochrones \label{tab:polynomials}}
\tablehead{
 \colhead{Association} &
 \colhead{\# stars in fit} &
 \colhead{min $V-K$} &
 \colhead{max $V-K$} &
 \colhead{c$_0$} & 
 \colhead{c$_1$} &
 \colhead{c$_2$} &
 \colhead{c$_3$} &
 \colhead{c$_4$} &
 \colhead{c$_5$}
}
\startdata
$\beta$~Pic    & 32 & 0.0 & 6.3 &  1.6265 &  2.1789 & -0.6631 &  0.2978 & -0.0550 &  0.00370 \\
TW~Hya         & 16 & 0.0 & 8.3 &  1.4555 &  0.8992 &  1.2706 & -0.5648 &  0.0915 & -0.00485 \\
AB~Dor         & 41 & 1.2 & 5.1 & -5.7448 & 13.6063 & -6.5304 &  1.6629 & -0.1949 &  0.00845 \\
Tuc-Hor        & 31 &-0.6 & 3.9 &  1.3926 &  1.8747 & -0.0707 &  0.4269 & -0.2199 &  0.02960 \\
$\epsilon$ Cha & 10 &-0.3 & 6.6 &  0.4050 &  1.2320 &  2.2331 & -1.2254 &  0.2256 & -0.01376 \\
\enddata

\tablecomments{Coefficients are for the equation
M$_V$=c$_0$+c$_1$($V-K$)+c$_2$($V-K$)$^2$+c$_3$($V-K$)$^3$+c$_4$($V-K$)$^4$+c$_5$($V-K$)$^5$}
\end{deluxetable}

\end{document}